\documentclass{PoS}

\usepackage{multirow}
\usepackage{setspace}
\usepackage{amsmath}
\newcommand{\bc}{\begin{center}}
\newcommand{\ec}{\end{center}}
\newcommand{\bi}{\begin{itemize}}
\newcommand{\ei}{\end{itemize}}
\newcommand{\bfig}{\begin{figure}[hbt]}
\newcommand{\efig}{\end{figure}}
\newcommand{\btab}{\begin{table}[hbt]}
\newcommand{\etab}{\end{table}}
\newcommand{\fm}{\mbox{fm}}
\newcommand{\eq}[1]{(\ref{eqn:#1})}
\newcommand{\Eqv}[1]{Eq. (\ref{eqn:#1})}
\newcommand{\fig}[1]{figure \ref{fig:#1}}
\newcommand{\Fig}[1]{Figure \ref{fig:#1}}
\newcommand{\cO}{\mathcal{O}}
\newcommand{\be}{\begin{equation}}
\newcommand{\ee}{\end{equation}}
\def\H{{\cal H}}

\title{ Status of dynamical ensemble generation}
\ShortTitle { Status of dynamical ensemble generation}

\author{\speaker{Chulwoo Jung}
\\
        Department of Physics, Brookhaven National Laboratory, Upton, NY 11973, U.S.A.\\
        E-mail: \email{chulwoo@bnl.gov}}

\abstract{ I give an overview of current and future plans of dynamical
QCD ensemble generation activities.
A comparison of simulation cost between different discretizations is made.
Recent developments in techniques and algorithms used in QCD dynamical
simulations, especially mass reweighting,  are also discussed.}
          
\FullConference{The XXVII International Symposium on Lattice Field Theory - LAT2009\\
		 July 26-31 2009\\
		 Peking University, Beijing, China}

\begin{document}

\section{Introduction}
Recently there has been a
remarkable progress in reaching continuum limit via Lattice QCD, made possible
by  better simulation algorithms
and better lattice discretizations to suppress lattice
spacing errors in moderate lattice spacings, in addition to steady advances in
computing hardware.
In many ways, it is more than just a rhetoric to say we are at the cusp of
generating realistic lattice QCD configurations.

Since 
an extensive review of systematic errors, theoretical and
practical issues of each lattice discretization schemes was given
at the last year's conference~\cite{Jansen:2008vs}, here I will 
focus more on a survey of ongoing lattice ensemble activities and technical details 
from different groups presented at the conference and updates since the last
year. More comprehensive review of physical quantities from these lattices will be made
elsewhere, for example in~\cite{Scholz:2009yz,Aubin:2009yh,VandeWater:2009uc,Lat09_Lubicz}.  
Also, a brief and incomplete review of recent progress and trends in dynamical
simulation algorithms is given with more details on mass reweighting technique.

It should be noted that while the list of dynamical lattice QCD ensemble 
generation activities presented here is extensive, it is limited to only
zero temperature ($T \geq L$) and "QCD-like" ensemble simulations, with 2 light
quarks. There are significant amount of ensemble generation activities for QCD
thermodynamics studies and beyond the standard model studies, especially in anticipation
of upcoming results from Large Hadron Collider(LHC). These topics are covered
in~\cite{Laine:2009ik} and \cite{Lat09_Pallante} respectively.

The organization is as follows. Section \ref{section:action} gives descriptions and
summaries of ongoing ensemble generating activities. A 
comparison of simulation cost for various lattice discretizations is  given in section
\ref{section:cost}. Section \ref{section:autocorrelation} summarizes available
measurements of autocorrelations, especially topological charge.
Summary of advances in simulation algorithms and techniques is given in
section \ref{section:algorithm}.

\section{Ensembles}

\label{section:action}
\subsection{Domain Wall Fermion(DWF)}
\label{section:DWF}
\begin{figure}[hbt]
\hspace{-0.05\textwidth}
\begin{minipage}{0.5\textwidth}
\vspace{-0.05\textwidth}
\bc
\includegraphics[angle=0,width=1.0\textwidth]{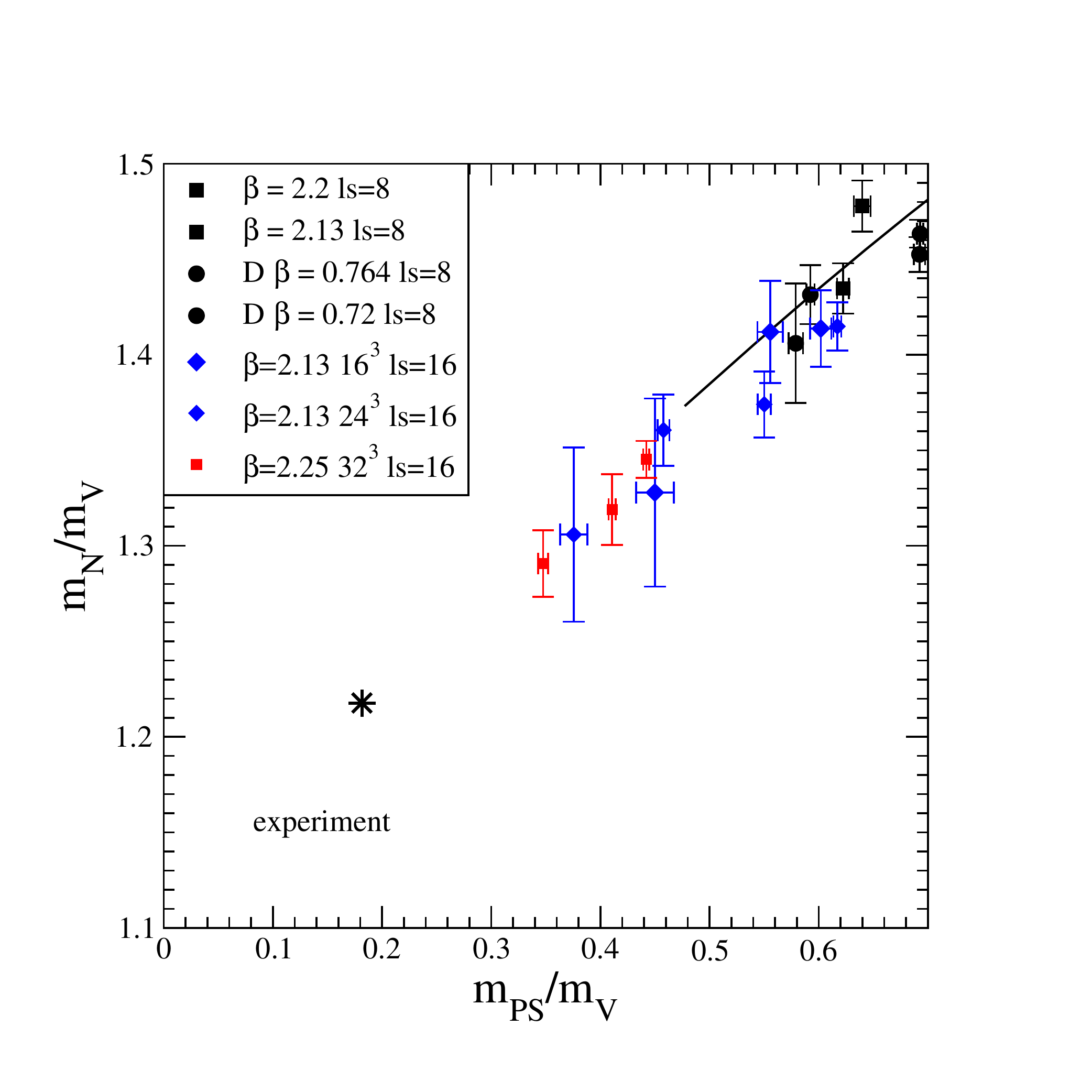}
\ec
\end{minipage}
\hspace{-0.00\textwidth}
\begin{minipage}{0.5\textwidth}
\vspace{-0.07\textwidth}
\begin{center}
\includegraphics[angle=0,width=1.1\textwidth]{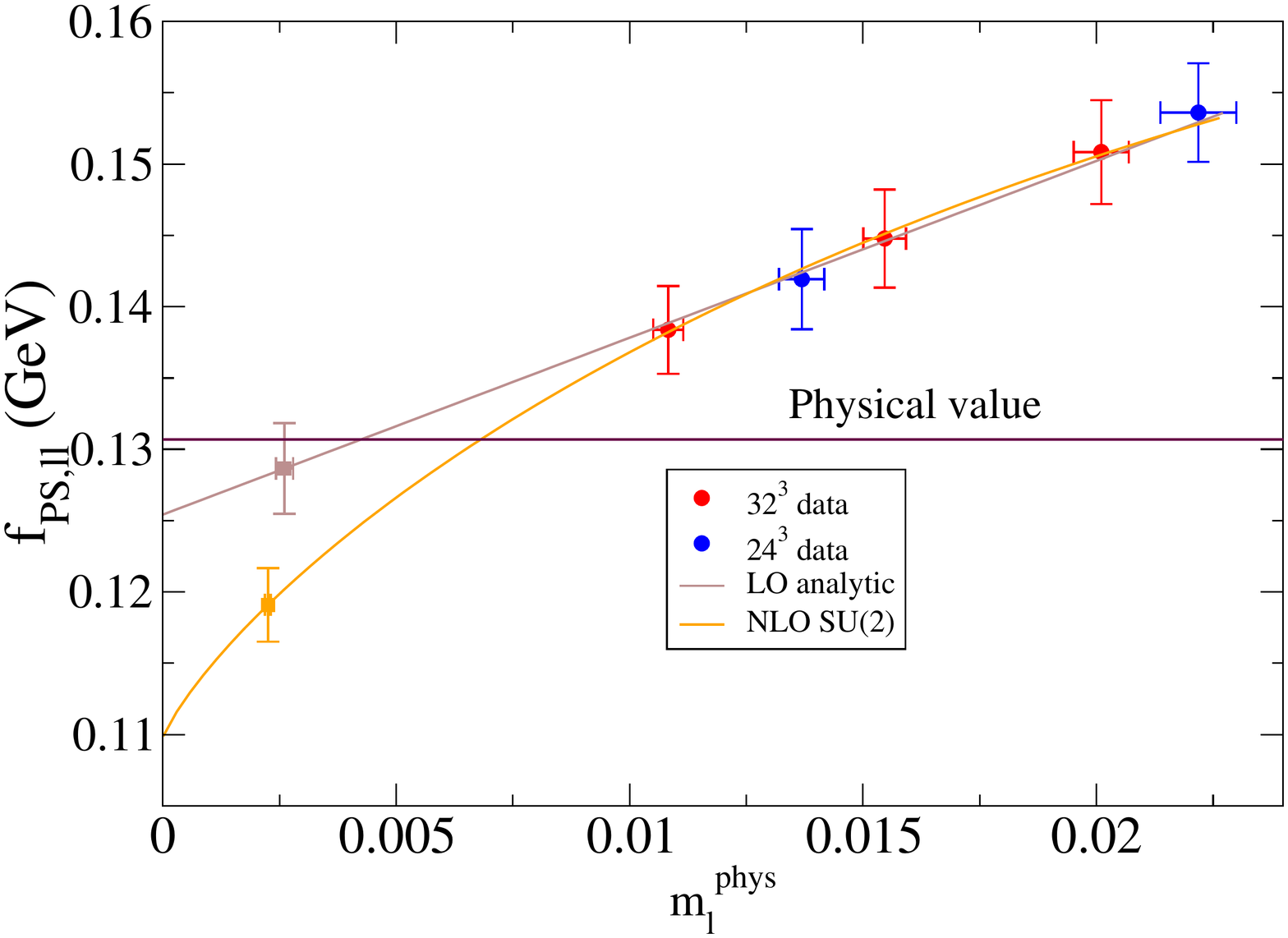}
\end{center}
\end{minipage}
\caption{Edinburgh plot for dynamical DWF lattices, from\cite{Lat09_Maynard}. }
\label{fig:DWF_Edin}
\vspace{-0.02\textwidth}
\caption{ Pseudoscalar decay constant($f_\pi$) with $m_{sea} = m_{val}$ for 
$24^3$ and $32^3$ DWF lattices, from\cite{Kelly:2009fp}.}
\label{fig:DWF_scaling}
\end{figure}
RBC/UKQCD collaborations have been generating $N_f=2+1$ DWF configurations
\cite{Allton:2008pn,Allton:2007hx,Antonio:2007pb} using 
DWF formalism from \cite{Furman:1994ky} and Iwasaki gauge
action\cite{Iwasaki:1983ck} which gives a good balance between preservation of
chiral symmetry and practicality for the lattice spacings studied so far. 
Iwasaki action was chosen instead of DBW2 action~\cite{Takaishi:1996xj}, which was
used for previous DWF studies, to ensure sufficient decorrelation in topological charge.

A recent deployment of IBM BG/P machines at Argonne Leadership Computing
Facility(ALCF) made the ensembles at the second lattice spacing much earlier
than initially anticipated. 
Currently ensembles in 2 lattice spacings are available:

\bi
\item $\beta=2.13, a  \sim 0.121 \fm, 
16^3\times 32 \times 16, m_\pi \sim 400, 526, 627$ Mev
\item $\beta=2.13, a  \sim 0.114 \fm,~24^3\times 64 \times 16$, 
$m^{\bar{MS}}_{res} \sim $ 8.5 Mev, $ m_\pi  \sim$ 328,~417 Mev
\item $\beta=2.25, a \sim 0.084 \fm,~32^4\times 64 \times 16$, 
$m^{\bar{MS}}_{res}\sim$ 2.45Mev, $m_\pi \sim$ 295,~350,~397 Mev
\ei

Figures \ref{fig:DWF_Edin} and \ref{fig:DWF_scaling} Shows the scaling
behavior between DWF ensembles with 2 different lattice spacings. 
The quantities measured within the
range of sea quark mass shows the scaling violation is within 2\%. This allows 
fitting both ensembles to NLO SU(2) ChPT in $a^2$ and $m_l$, where only leading
order(LO) ChPT low energy constants (LEC's)
have $a$ dependence. Reweighting in dynamical strange quark mass
(section \ref{section:str_rw})
 and interpolation in valence mass is used to reach the physical strange quark
mass point.  Numbers in physical units for the last 2 sets of ensembles are
from this SU(2) ChPT global fit using both ensembles. Details of the fitting
procedure and the results for  pseudoscalar meson and decay constants is
reported in \cite{Mawhinney:2009jy}. 
$B_K$ and detailed scaling study is reported
in  \cite{Kelly:2009fp}. Results on hadron masses are in \cite{Lat09_Maynard}.

{
\subsection{DWF with Auxiliary Determinant (Modified Gap DWF)}
\begin{figure}[hbt]
\hspace{-0.05\textwidth}
\begin{minipage}{0.5\textwidth}
\vspace{-0.1\textwidth}
\includegraphics[angle=0,width=1.2\textwidth]{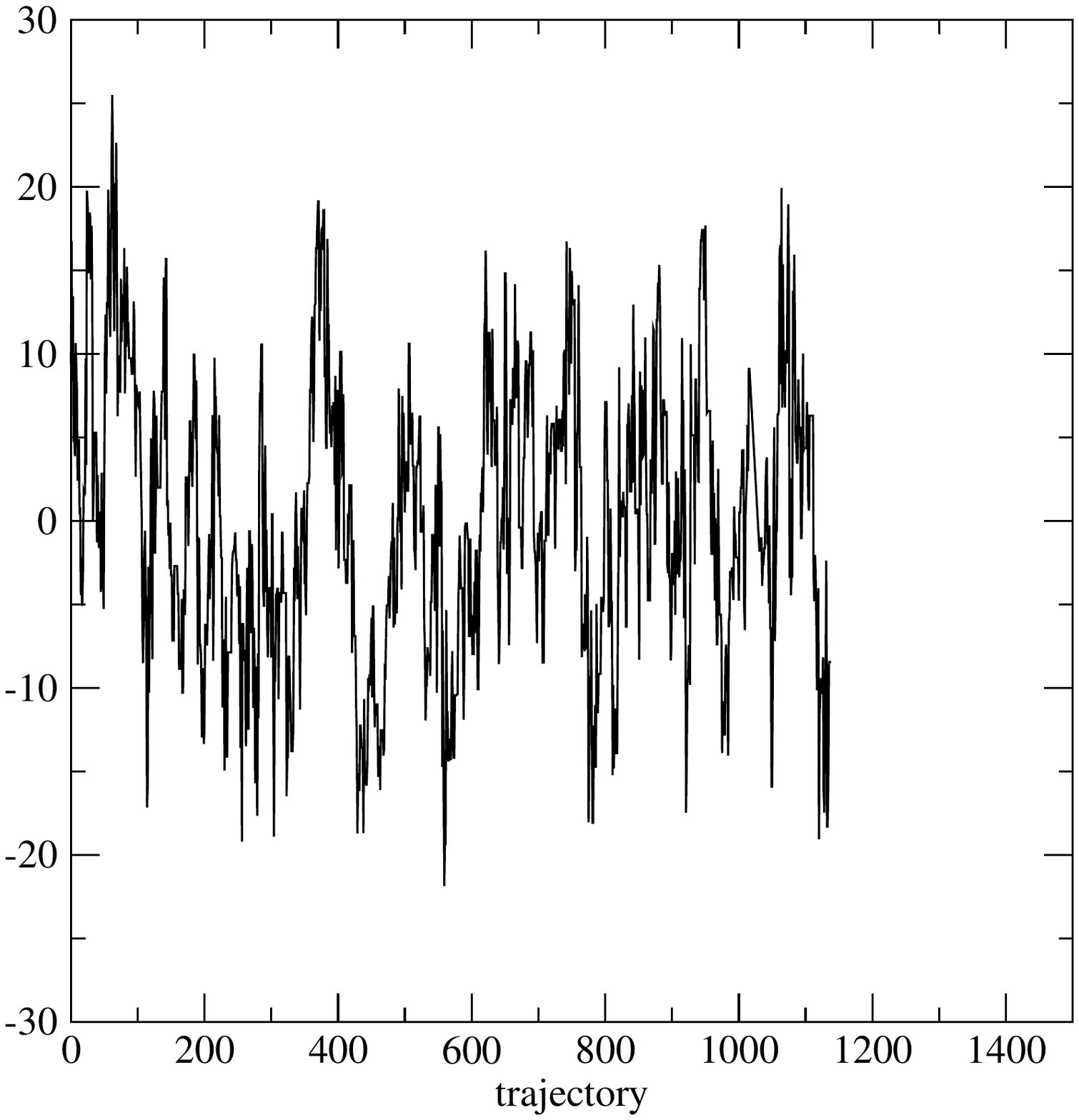} 
\end{minipage}
\hspace{-0.02\textwidth}
\begin{minipage}{0.5\textwidth}
\includegraphics[angle=0,width=\textwidth]{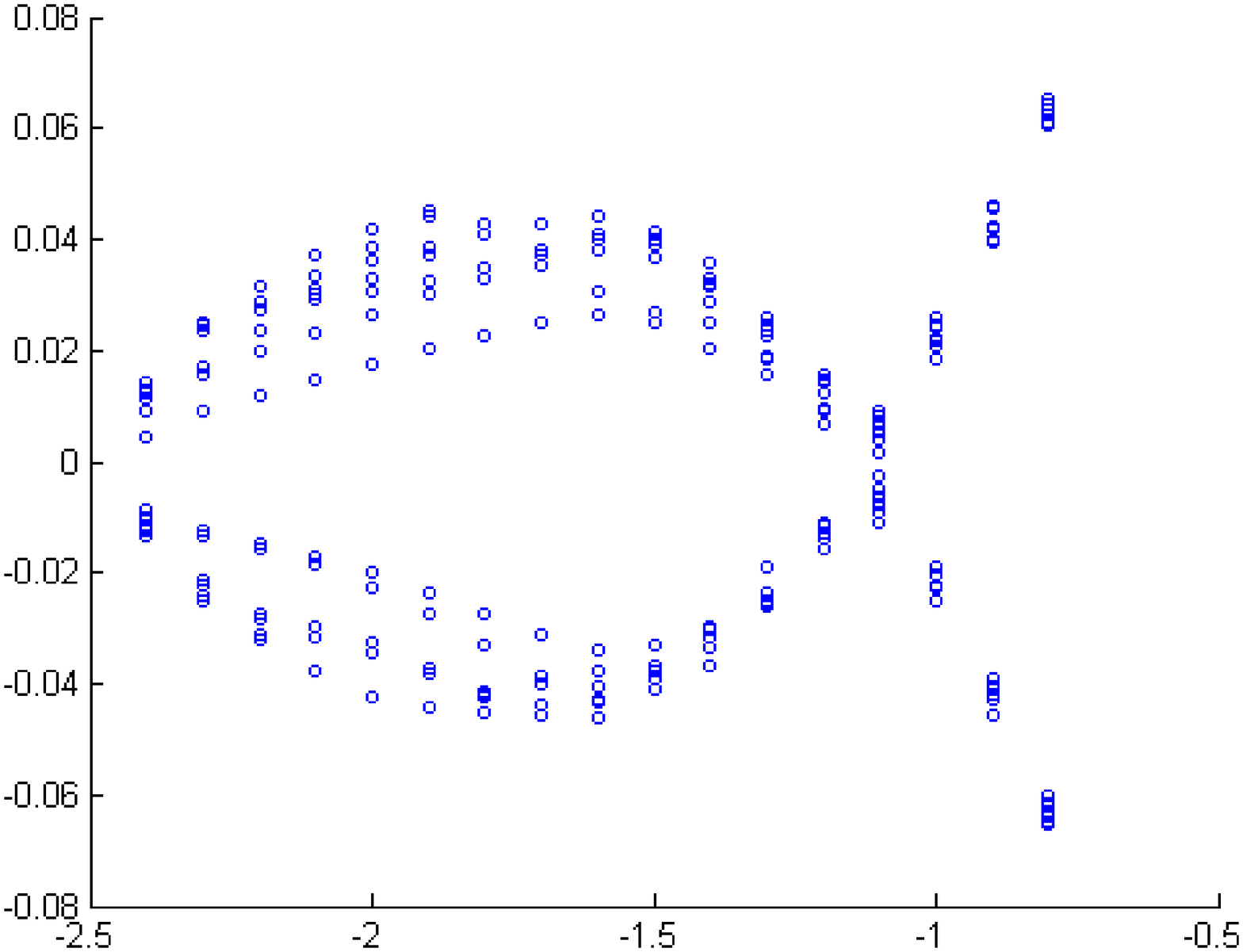}
\end{minipage}
\caption{Time history for $\beta=1.75, m_l=0.042, 32^3 \times 64\times 32 $ ensemble}
\vspace{-0.02\textwidth}
\caption{Eigenvalue flow diagram for a $\beta=1.75, 16^3\times 8 \times 32 $ lattice}
\label{fig:AuxDet_top}
\end{figure}

Lattice dislocations which induce residual chiral symmetry breaking in DWF
is currently the biggest obstacle for DWF and overlap fermions at larger
lattice spacings. ($a \geq 0.1$ fm). Hence,
controlling residual mass at these lattice spacings
is crucial for DWF studies of quantities which requires large
lattice volume, such as QCD thermodynamics, nucleon matrix elements,
weak matrix elements via  direct studies of $K \rightarrow \pi \pi$ process on
the lattice. 

Various approaches have been used to suppress dislocations in the past. 
Different gauge actions such as DBW2 action\cite{Takaishi:1996xj} have 
shown to suppress dislocations successfully, but at the expense of
suppression of topology tunneling at smaller lattice spacings.
Alternatively, additional fermion action can be used to suppress 
dislocations, as they are  related to near zero modes of Wilson Dirac
operators. This idea was first introduced in \cite{Vranas:2006zk} and later
used in dynamical overlap simulation by JLQCD collaboration\cite{Aoki:2008tq},
where a ratio of Wilson fermion determinants were used instead.
It should be noted that while formally fermions are added, these are not
related to physical quarks and this is effectively just changing gauge action
after they are integrated out.

}

For the lattice spacings currently being investigated, determinants used in
either \cite{Vranas:2006zk} or \cite{Aoki:2008tq} appear to suppresses the
topology tunnelling too strongly. 
To circumvent this, a small imaginary mass is also added to the numerator to
ensure the zero eigenvalues are not completely suppressed. 

Supression factor from these additional terms is given by
{
\begin{eqnarray}
\mathcal{W}(M_5, \epsilon_f, \epsilon_b) =
\frac{\det\left[D_{W}(-M_5+\imath\epsilon_f\gamma^5)^\dag
D_{W}(-M_5+\imath\epsilon_f\gamma^5)\right]}{\det\left[D_{W}(-M_5+\imath\epsilon_b\gamma^5)^\dag
D_{W}(-M_5+\imath\epsilon_b\gamma^5)\right]} & \nonumber \\
= \frac{\det\left[H_{W}(-M_5)^\dag
H_{W}(-M_5)+\epsilon_f^2\right]}{\det\left[H_{W}(-M_5)^\dag
H_{W}(-M_5)+\epsilon_b^2\right]}
= \prod_i \frac{\lambda^2_i+\epsilon^2_f}{\lambda^2_i+\epsilon^2_b} &
\label{eqn:AuxDet}
\end{eqnarray}

Where $\lambda_i$ are eigenvalues of Hermitian Wilson dirac operator with mass
$-M_5, H_W(-M_5) = \gamma_5 D_W(-M_5)$.
Eq.~\eq{AuxDet} gives 
$\sim 1$ for $\lambda_i \gg \epsilon_b, \epsilon_f$, while $\sim \epsilon^2_f/
\epsilon^2_b$ for $\lambda_i \ll \epsilon_f, \epsilon_b$.
}
$\epsilon_f=0$ corresponds to what is used in~\cite{Aoki:2008tq} and only
numerator with $\epsilon_f=0$ was used in~\cite{Vranas:2006zk}.
{

\begin{table}[hbt]
\bc
\begin{tabular}{c|c|c|c|c|c|c}
	\hline\hline
	\multicolumn{7}{c}{Aux. Det. $\beta=1.75, a \sim 1.4 \mbox{Gev},
\epsilon_f/\epsilon_b = 0.02/0.5,  a m_{res} \sim 0.0019$ } \\
	\hline
  		$L/a$ & $m_sa$ & $m_la$ & $L(fm)$  & $ m_{PS}$(Mev) & $\tau$(MD) & Accept.\\
	\hline
	$32^3 \times 64 \times 32$ & 0.045 & 0.0042 & $\sim$ 4.5 & $\sim 250$ & $\sim$ 1200 & $\sim$ 70\% \\
	$32^3 \times 64 \times 32$ & 0.045 & 0.001 & $\sim$ 4.5 & $\sim 200$ &  $\sim$ 250& $\sim$ 70 \% \\
\end{tabular}
\ec
\caption{List of ongoing ensemble generation using DWF with \protect\Eqv{AuxDet}.}
\label{table:AuxDet}
\end{table}

Table \ref{table:AuxDet} shows the ongoing DWF simulations with the auxiliary
determinant.
Although it turned out it is necessary to use a rather large $\epsilon_b$ to make enough
suppression of the residual mass, which causes the shifts in gauge coupling 
$\beta$, a factor of 5-7 decrease in residual mass was observed after the
scales are matched by locating transition temperature~\cite{Renfrew:2009wu}.

\Fig{AuxDet_top} shows the topological charge evolution of these
ensembles and Wilson Dirac operator eigenvalue flow diagram for an ensemble
with same lattice spacing and smaller volume, which suggests the
topology is being sampled well while dislocations which causes the small
eigenvalues near $-M_5= 1.8$ are suppressed as intended. 

Preliminary results form $m_\pi \sim 250$Mev ensembles suggests that the
scaling error between $a \sim 0.14$fm AuxDet ensembles and existing DWF
ensembles is at a few percent level, and it is possible to fit
DWF ensembles with and without the auxiliary determinant in a fashion in \cite{Mawhinney:2009jy}, by
allowing different $a^2$ dependence to LO LEC's in the ChPT fitting. 
Result of this analysis will be forthcoming.

\subsection{$a^2$, tadpole improved staggered action (Asqtad)}
\hspace{-0.1\textwidth}
\vspace{-0.1\textwidth}
\begin{figure}[hbt]
\begin{minipage}{0.5\textwidth}
\vspace{-0.3\textwidth}
\includegraphics[angle=0,width=1.1\textwidth]{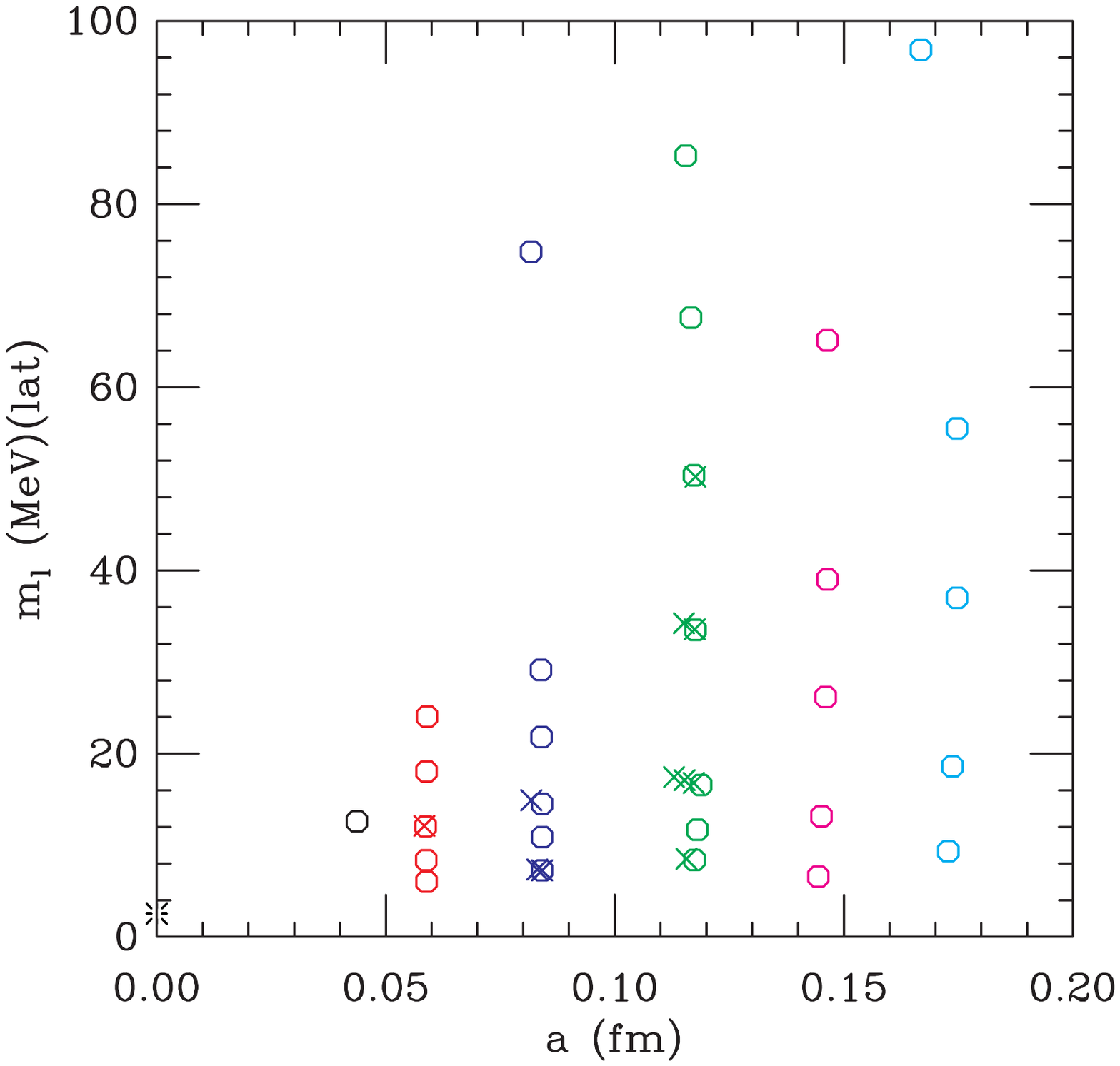} 
\vspace{-0.3\textwidth}
\end{minipage}
\begin{minipage}{0.5\textwidth}
\vspace{-0.05\textwidth}
\includegraphics[angle=0,width=\textwidth]{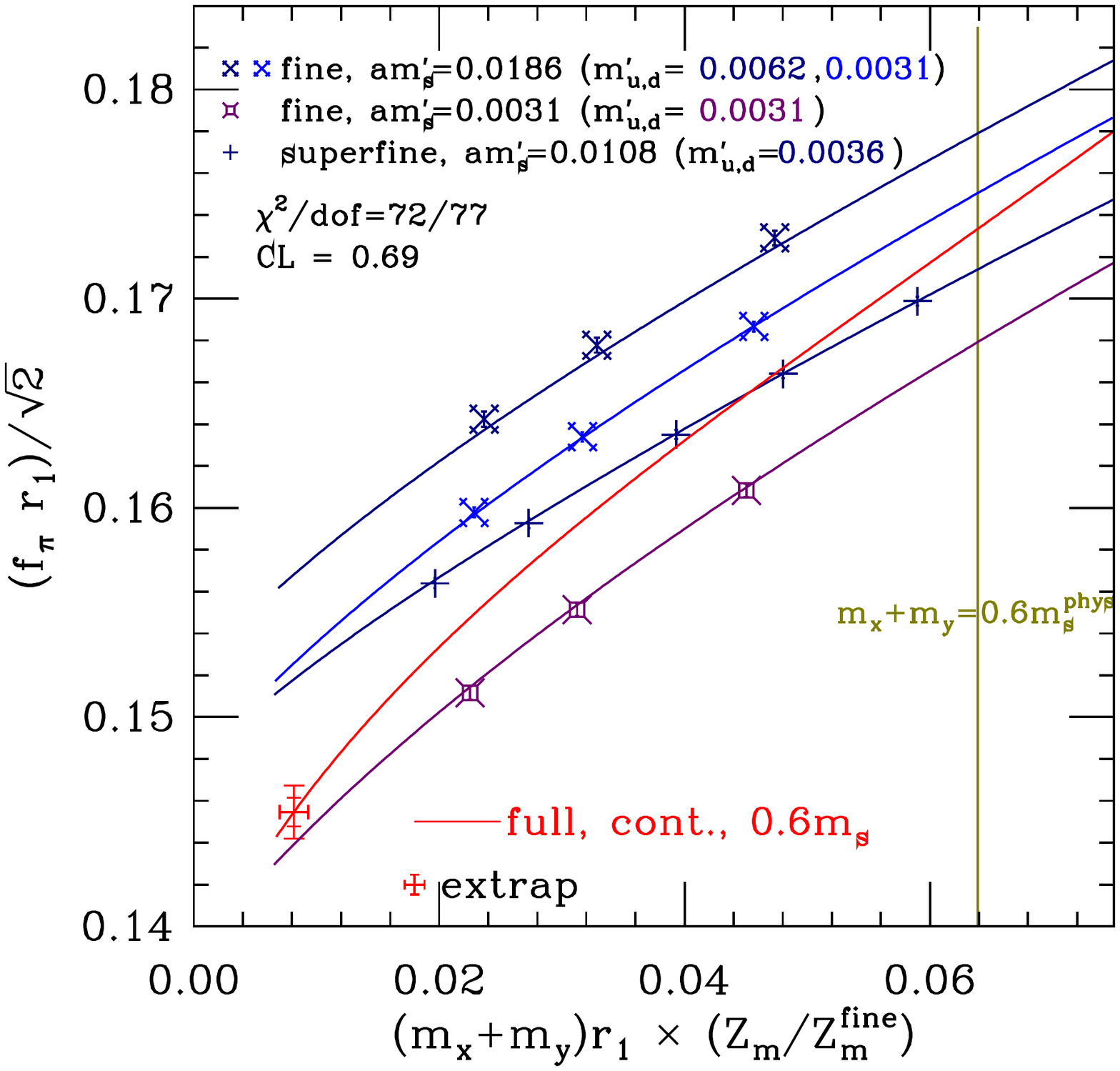}
\end{minipage}
\vspace{-0.1\textwidth}
\caption{
Lattice spacing and light quark masses of available $N_f=2+1$ dynamical
Asqtad ensembles, from \cite{Bazavov:2009bb}.
}
\label{fig:milc_runs}
\vspace{-0.02\textwidth}
\caption{SU(3) ChPT fit to $f_\pi$  from \cite{Bazavov:2009tw}.
Red line shows the continuum extrapolation with $m_s \sim 0.6 m_s^{phys}$.
} 
\label{fig:milc_ps}
\end{figure}

MILC collaboration has been generating $N_f=2+1$ dynamical ensembles with 
improved staggered fermion action(Asqtad)\cite{Orginos:1999cr}, designed to
suppress taste symmetry violation present in staggered fermions, with multiple
lattice spacings and quark masses. An extensive review of ensembles and
physics results is given in \cite{Bazavov:2009bb}. 

As shown in \fig{milc_runs}, most of the simulation is done $0.4 \geq m_l/m_s
\geq $0.1, which gives $m_{\pi} \geq  220$~Mev, at multiple lattice spacings,
$a \sim$ 0.15~(usually referred as "extra-coarse"), 0.12~(coarse), 0.09~(fine),
0.06~(superfine), 0.045~(ultrafine) fm.
 In addition to these, there are ensembles generated  at $m_s$ near 60\% of 
physical strange quark mass, to aid SU(3) ChPT studies.
Most recent continuum extrapolation of Sommer scale $r_1$, set by $f_K$, gives
0.3117 fm~\cite{Bazavov:2009tw}.
Typically staggered chiral perturbation theory\cite{Aubin:2003mg,Aubin:2003uc}
is used for continuum extrapolations.
An update of SU(3) ChPT fitting results are reported in \cite{Bazavov:2009tw}. 

It should also be noted that MILC ensembles have been extensively used for various mixed
action studies, for example \cite{Aubin:2009jh}, where valence quarks with
different discretizations are used, due to
early availability of ensembles with multiple lattice spacings and quark
masses. For these studies, the effects of taste symmetry breaking also has to
be accounted for and this is typically done via mixed action chiral
perturbation theories such as\cite{Bar:2005tu}.

\subsection{Highly Improved Staggered Quarks (HISQ)}
\label{section:HISQ}
\begin{figure}[hbt]
\hspace{-0.05\textwidth}
\begin{minipage}{0.51\textwidth}
\vspace{0.00\textwidth}
\includegraphics[angle=0,width=\textwidth]{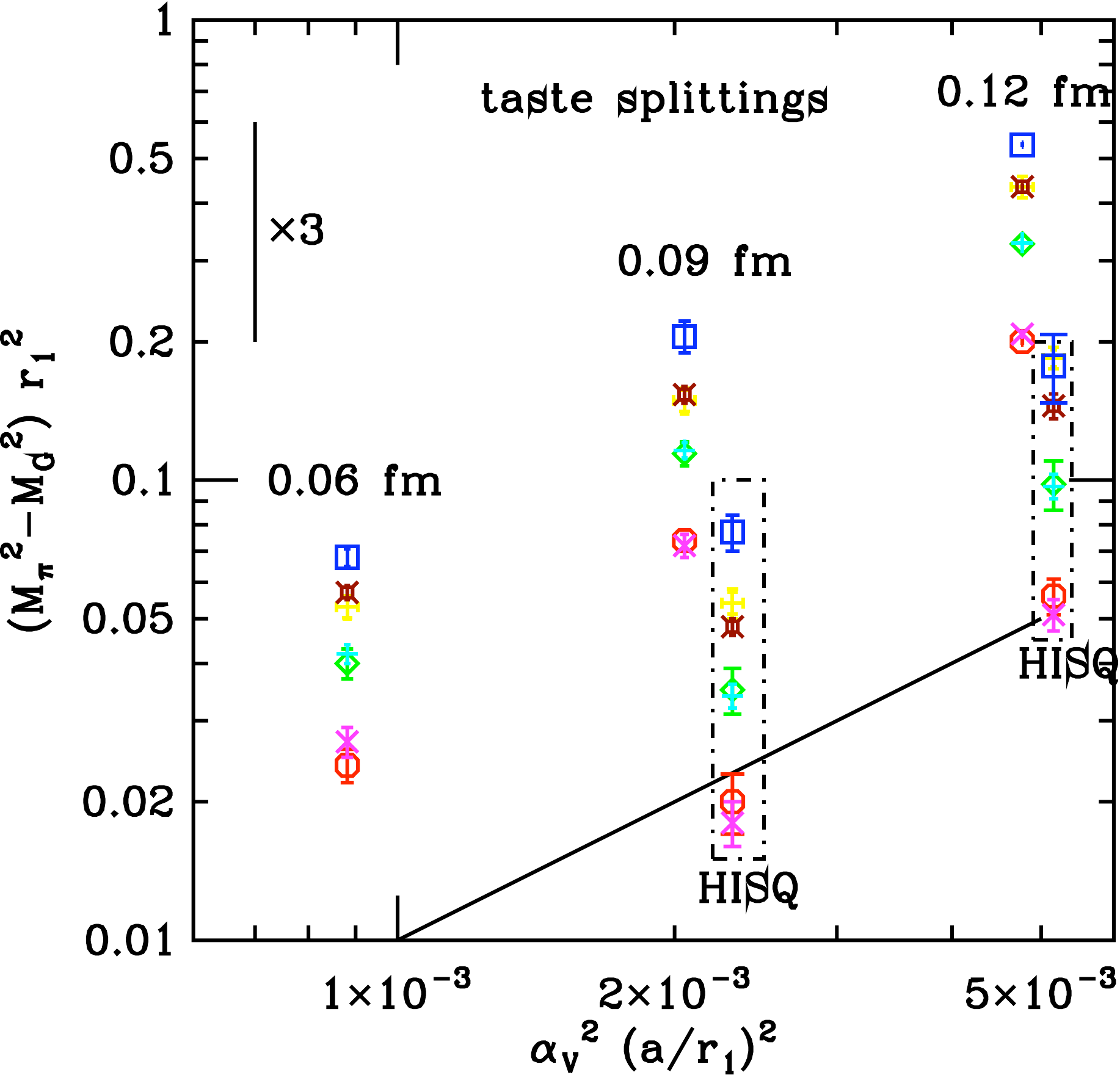}
\end{minipage}
\hspace{-0.00\textwidth}
\begin{minipage}{0.49\textwidth}
\vspace{-0.00\textwidth}
\includegraphics[angle=0,width=\textwidth] {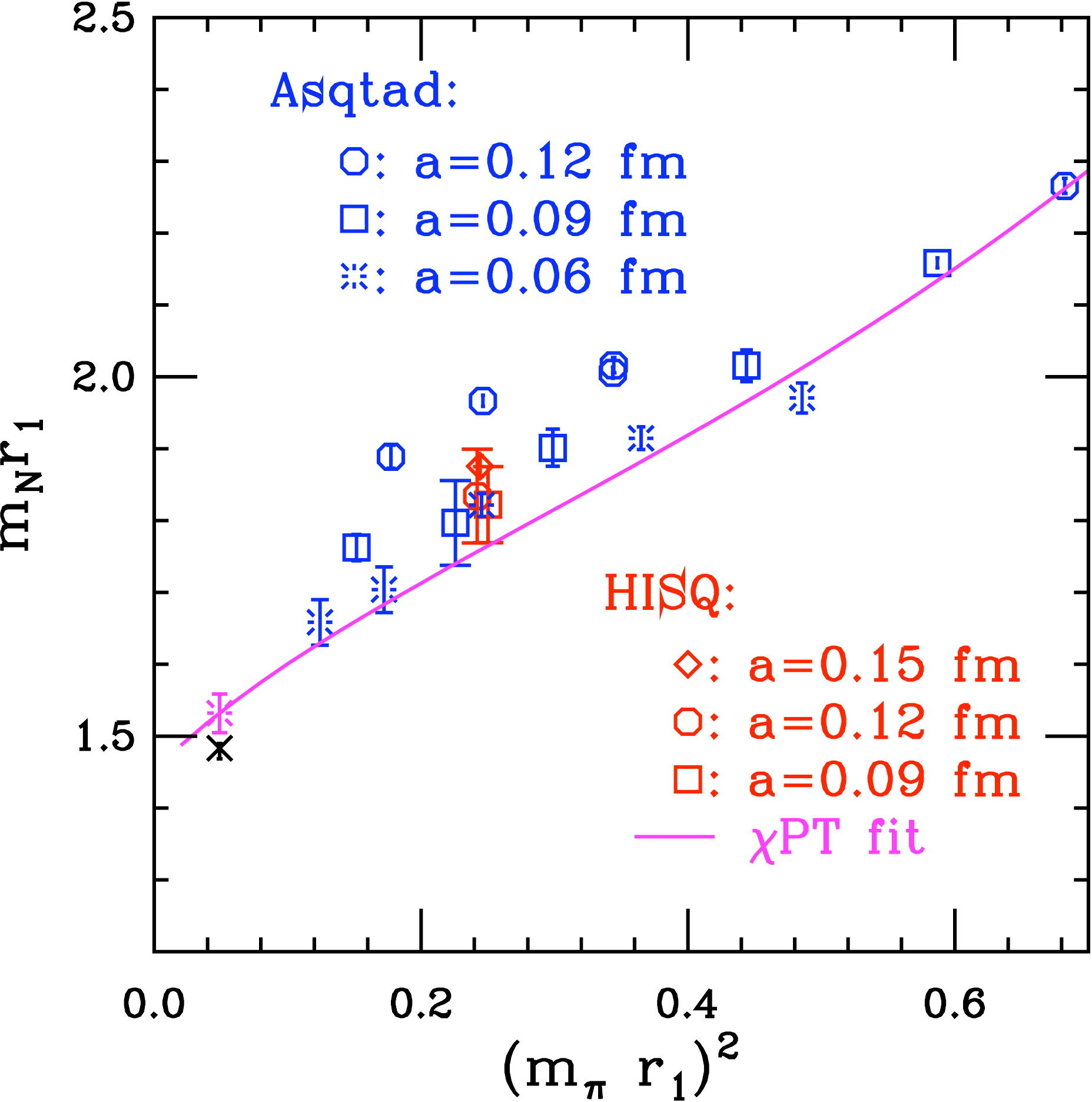}
\end{minipage}
\caption {Taste splitting of Asqtad and HISQ action, measured by the difference of squared
pseudoscalar meson masses with different tastes\cite{Bazavov:2009wm}. $M_I$ and $M_G$ is the mass
of taste singlet and Goldstone pseudoscalar respectively.
}
\label{fig:hisq_splitting}
\caption{ Scaling plot of nucleon mass for Asqtad and HISQ actions\cite{Bazavov:2009wm}.} 
\end{figure}

HISQ action \cite{Follana:2006rc} is a staggered fermion action, improved further from
Asqtad action by introducing additional Fat-7 smearing followed by projection
to U(3) using Cayley-Hamilton theorem, used also
in~\cite{Morningstar:2003gk,Hasenfratz:2007rf}, before 
combined in a similar fashion to Asqtad action.

Preliminary studies of dynamical ensembles generated with HISQ action
indicates  that while it is about 2 times more expensive per MD units
compared to those of Asqtad actions at the similar lattice spacing,
the mass splitting between pseudoscalar mesons with
different tastes are reduced 
by a factor of $2.5\sim 3$ 
(Fig. \ref{fig:hisq_splitting}).

While the U(3) projection after additional smearing have shown to be effective
in suppressing taste symmetry breakings,  
 this also causes the fermion force for MD steps to be large when the smeared
link happens to have a small eigenvalue, resulting in low acceptance.
These difficulty is avoided by replacing
$Q^{-1/2}$ with   $(Q+\delta I)^{-1/2} (\delta \sim 5 \times 10^{-5})$ during
moleculardynamics~(MD) steps, where $Q$ is a smeared link to be projected to
U(3).  While this effectively changes the hamiltonian for MD step, 
Metropolis step after each trajectory corrects for the discrepancy.
Also, dynamical charm quark can be included with $O((am_c)^4))$ errors removed,
but the coefficient for a straight 3-link term (Naik term) should be set different for different quarks.

Preliminary studies with HISQ actions for both thermodynamic studies ($N_f=2+1$)
\cite{Lat09_Bazavov} and zero temperature studies with dynamical charm quark 
($N_f=2+1+1$)\cite{Bazavov:2009wm} are reported at the conference.

\subsection{$\cO(a)$ improved Wilson (Clover) action}
\label{section:Clover}
There are ensemble generation activities by several collaborations using
Clover actions: Budapest-Marseilles-Wuppertal(BMW)
collaboration, PACS-CS collaboration, QCDSF collaboration
and Coordinated Lattice Simulations(CLS) collaboration.  All activities are
aimed at generating ensembles with the pion mass close or at the physical
value.
The main features of simulations are summarized in table \ref{table:Clover}.
\begin{table}[hbt]
\begin{tabular}{c|c|c|c|c|c}
\hline
Collaboration 	& BMW 			& PACS-CS &\multicolumn{2}{c|}{QCDSF } 	& CLS   \\
\hline
$N_f$		&	2+1		& 2+1	  & 2  		& 2+1(SLiNC)
& 2	\\
\hline
Gauge action	& Tree Symanzik	& Iwasaki & Wilson	& Tree Symanzik	 	&Wilson \\
\hline
$c_{sw}$	& Tree level 		& NP	  & \multicolumn{2}{c|}{NP}
& NP \\
\hline
Smearing	& Stout			&	  &		& Stout 		& \\
\hline
Algorithm	& (R)HMC		& DDHMC	  & \multicolumn{2}{c|}{(R)HMC}		& DDHMC\\ &			& +PHMC	  & \multicolumn{2}{c|}{}			& +deflation\\
\hline
$a$(fm)		& 0.065-0.125		&  0.09	  & 0.072-0.09	& $\sim$0.08			&0.04-0.08 \\
\hline
$m_\pi$(Mev)	& $> \sim$ 190		& 156-702 & 140 - 1010	& 			& 360 - 520\\
\hline
This conference&\cite{Lat09_Ramos} &\cite{Aoki:2009ix}
&\cite{Lat09_Schierholz}&\cite{Bietenholz:2009fi}&\cite{Capitani:2009tg}
\end{tabular}
\caption{Summary of dynamical Clover ensemble generation activities.}
\label{table:Clover}
\end{table}

\bfig
\begin{minipage}{0.5\textwidth}
\includegraphics[angle=0,width=\textwidth]{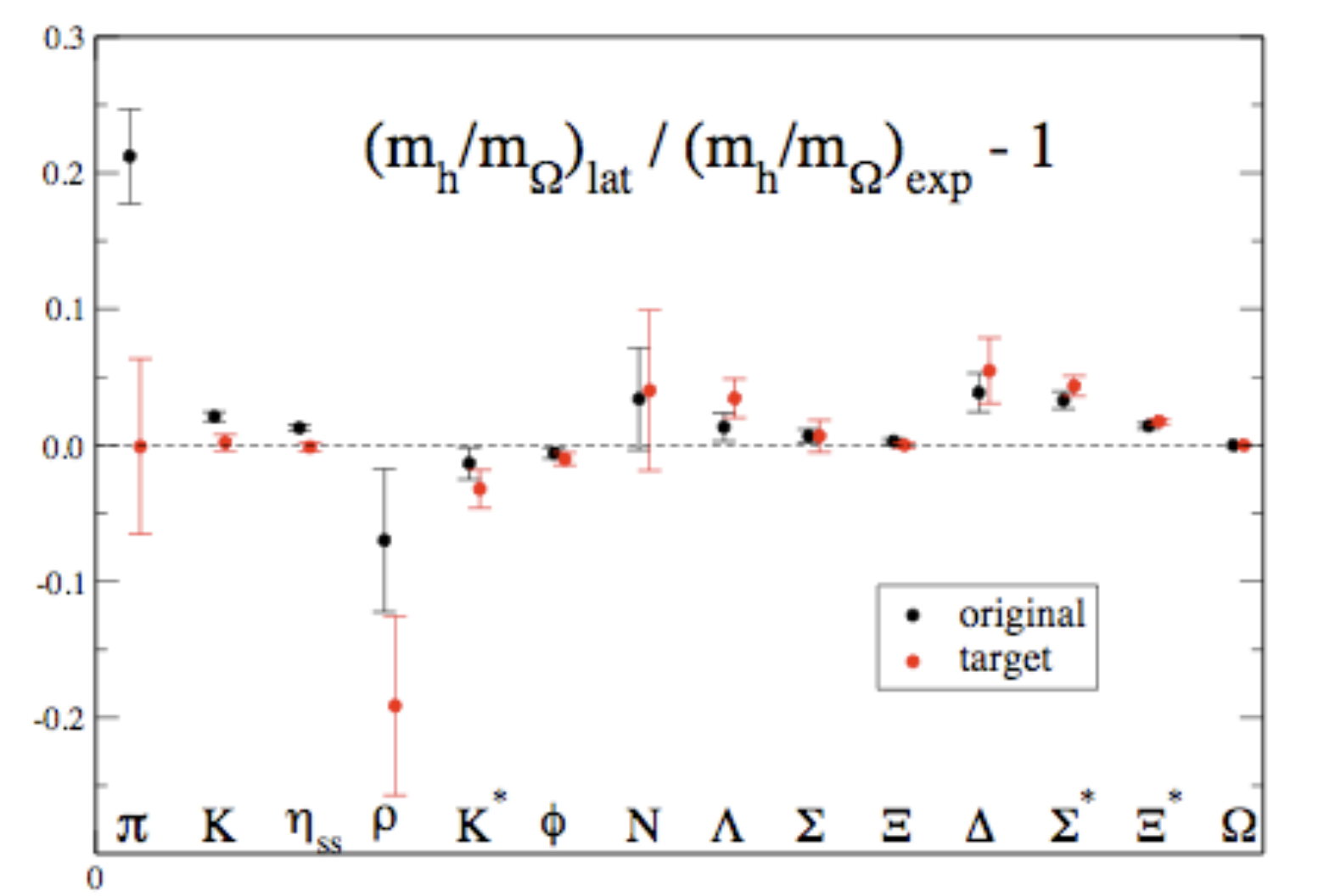}
\end{minipage}
\begin{minipage}{0.5\textwidth}
\includegraphics[angle=0,width=\textwidth]{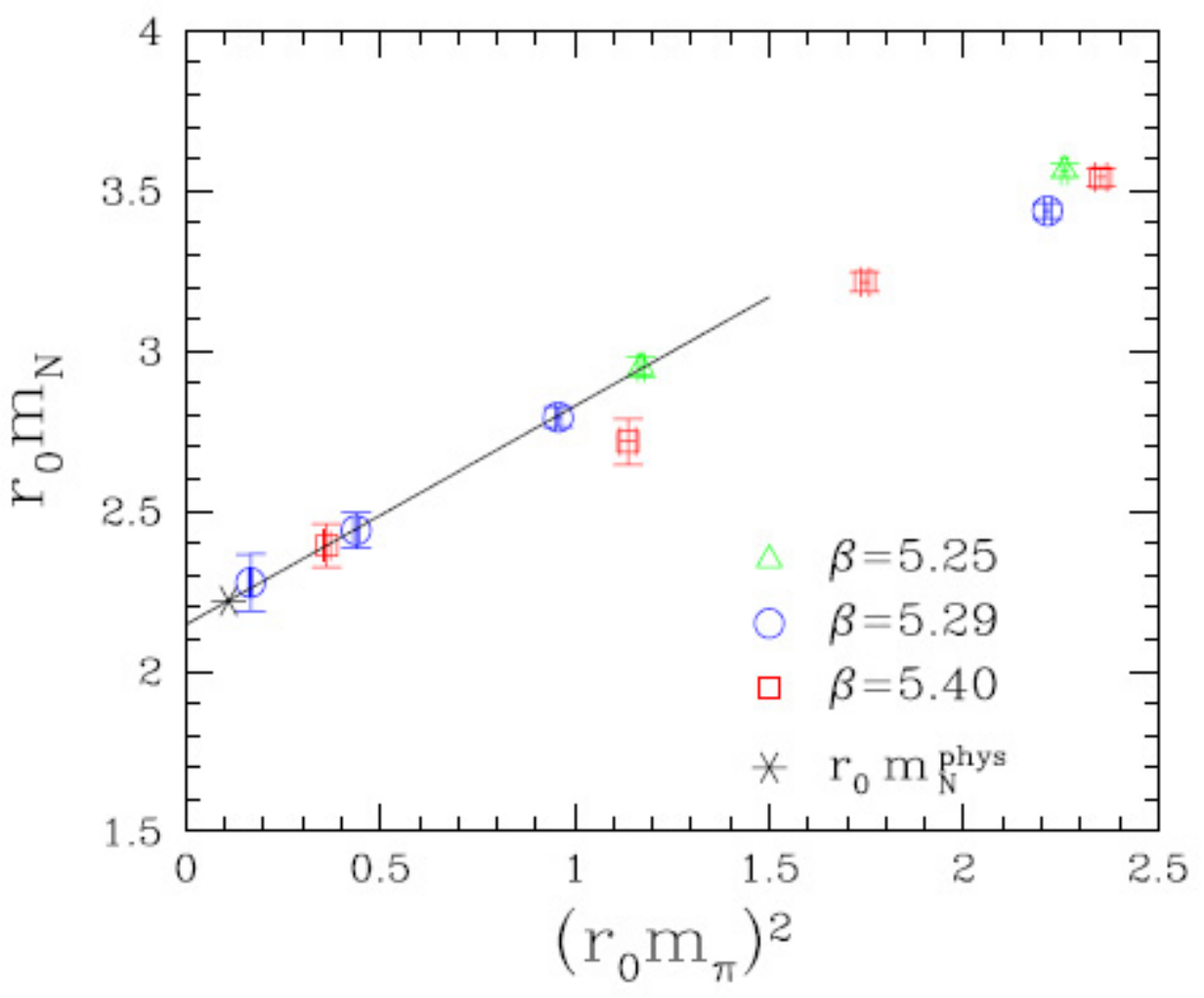}
\end{minipage}
\caption{ Deviation of hadron masses from physical values, normalized by $m_\Omega$,
from \cite{Aoki:2009ix}.}
\label{fig:PACS_mass} 
\caption{ Nucleon mass versus $m_\pi^2$ from QCDSF $N_F=2$ ensembles. $M_N \sim m_\pi^2$
behavior is observed.  $r_0=0.467$ fm is used.}
\efig

While nonpurtabatively determined $c_{sw}$ is used for most of simulations, ensembles
generated by BMW collaboration uses tree level value, but instead rely on up
to six levels  of stout smearing\cite{Morningstar:2003gk} to suppress lattice
discretization errors. The main results from
these ensembles were reported in\cite{Durr:2008zz,Durr:2008rw,Lellouch:2009fg} and update of
$f_K/f_\pi$ is given in \cite{Lat09_Ramos}.

PACS-CS\cite{Aoki:2008sm,Ishikawa:2009vc} has applied mass reweighting technique(section
\ref{section:reweighting}) to tune both light and strange dynamical masses to reach physical 
point for figure \ref{fig:PACS_mass}. 
Currently available configurations have a physical size $L \sim$ 3 fm, $m_\pi
L \sim  2.2$.  
 Simulation with $L \sim $6~fm is under way.

QCDSF collaboration has generated extensive $N_f=2$ ensembles and the results for hadron
masses from those configurations are given in \cite{Lat09_Schierholz}.  They are also
working on $N_f=2+1$ ensembles using stout smearing and
nonperturbative $c_{sw}$, called SLiNC action. A preliminary study for tuning quark
masses to physical point is described in \cite{Bietenholz:2009fi}. $r_0=0.5$ fm is used to set the scale.

CLS collaboration\cite{CLS_wiki}
aims to generate Clover ensembles in a wide range of 
lattice spacings, quark masses and lattice volumes using deflation accelerated DD-HMC\cite{Luscher:2007es}.
Preliminary results are reported in \cite{Capitani:2009tg}.

{
\subsection{Twisted mass Wilson(tmWilson)}
European Twisted Mass Collaboration(ETMC) has been generating 
both $N_f=2$ and $N_f=2+1+1$ dynamical tmWilson ensembles
\cite{Boucaud:2007uk,Boucaud:2008xu} with maximal twist. 
$N_f=2$ ensembles are generated with tree-level Symanzik gauge action, 
0.053 fm $ < a < 0.1$ fm, 280~Mev $ < m_\pi < $ 650Mev and   2.0 fm$ < L < 2.5$\fm.
\bfig
\hspace{-0.05\textwidth}
\begin{minipage}{0.5\textwidth}
\vspace{-0.1\textwidth}
\includegraphics[angle=0,width=\textwidth]{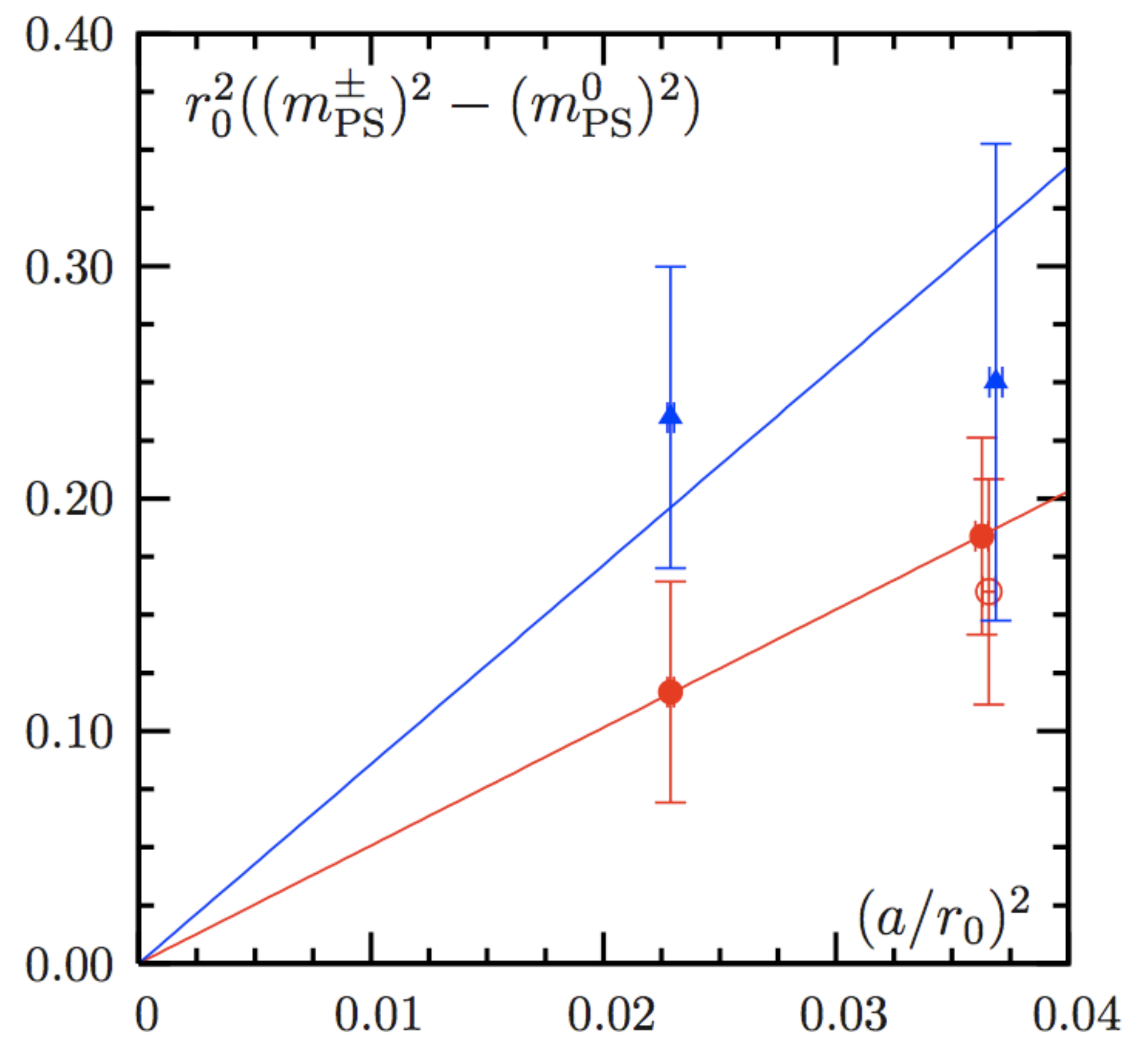}
\end{minipage}
\begin{minipage}{0.5\textwidth}
\includegraphics[angle=0,width=1.25\textwidth]{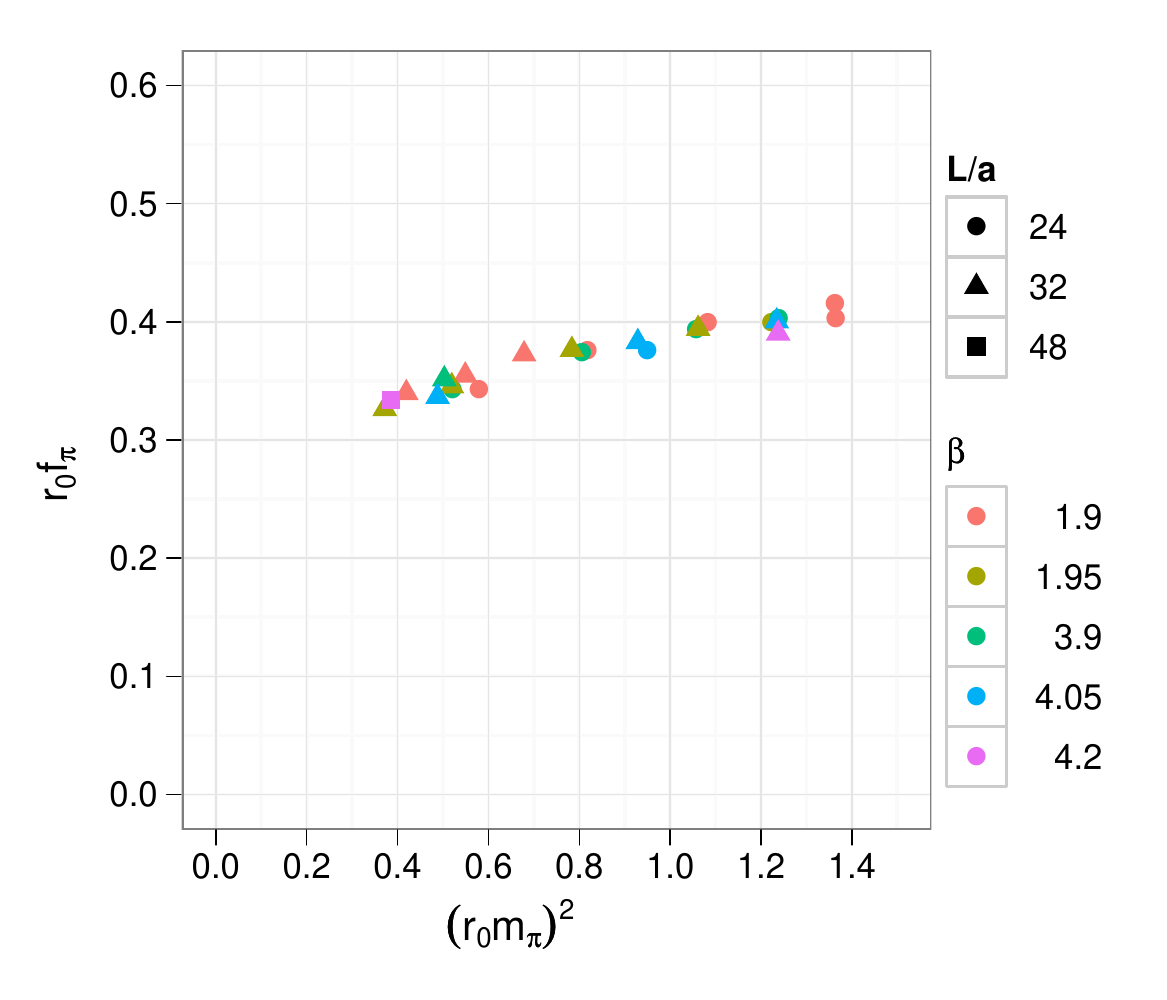}
\end{minipage}
\caption{
$(m_\pi^{\pm})^2- (m_\pi^0)^2$ vs. $a^2$ for tmWilson $N_f=2$ ensembles.}
\caption{ $f_\pi$ vs $m_\pi$ for tmWilson $N_f=2$ and 2+1+1 ensembles\protect\cite{Lat09_Reker}.}
\label{fig:ETMC_split}
\efig

A more detailed scaling study on $N_f=2$ ensembles shows a good scaling
between ensembles with 4 different lattice spacings for many quantities. An exception to this is the
mass splitting between charged and neutral pseudoscalar mesons from the breaking of
isospin symmetry at nonzero lattice spacing. \Fig{ETMC_split} shows while the behavior is consistent with $O(a^2)$ effect, it still is significant in the lattice spacings studied. This may be understood via
a Symanzik-type analysis\cite{Frezzotti:2007qv}.
More details of $N_f= 2$ ensembles is reported in \cite{Lat09_Herdoiza}. 

$N_f=2+1+1$ ensembles are generated with Iwasaki gauge action, at $a \sim 0.078,
0.086$fm, 280Mev, $ < m_\pi < $ 500Mev, $L < \sim 2.8$\fm. 
Polynomial HMC(PHMC)\cite{Chiarappa:2005mx} is used for the fermion simulation.
Twist parameters for heavy quarks, $\mu_\sigma$ and $\mu_\delta$, 
are fixed by tuning to  physical values of $K$ and $D$ mass while keeping $m_{PCAC,l}$
close to zero.
This is done without calculating all the possible excited states such as $K +
n\times \pi$ between K and D meson masses by assuming some
properties about our trial states and the corresponding correlation matrices.
it turned out $\mu_\delta$ was slightly underestimated for some of the ensembles. This is planned to be adjusted by
reweighting\cite{Baron:2008xa}.
More details of $N_f= 2+1+1$ ensembles is reported in \cite{Lat09_Reker}. 

In addition to this, $N_f=4$ ensemble is being generated for nonperturbative
renormalization with RI-MOM scheme\cite{Lat09_Lopez}. 

\subsection{Anisotropic Clover}
\label{section:AniClo}
Hadron Spectrum Collaboration(HSC) has been generating $N_f=2+1$ anisotropic clover
configurations at $ a_s=0.125~\fm, \xi = a_s/a_t = 3.5,  a_t \sim 5.6$~Gev, $L = 3 \sim 4$ fm, $m_\pi=$ 230, 360 Mev to
study various quantities such as resonance spectroscopy and nuclear forces.
Stout smearing is used on spatial links only
to suppress discretization errors further while preserving locality of transfer matrix. 
Strange quark is tuned  to
keep $s_\Omega = 9 (2m_K^2-m_\pi^2)/(2m_\Omega)^2$ constant while approaching continuum,
as shown in \fig{newport_news}. More details of algorithm and mass tuning are
given in \cite{Edwards:2008ja,Lin:2008pr}.
Results using these ensembles on 
excited nucleon spectroscopy \cite{Cohen:2009zk},
$\pi \pi$ states using distilled quark propagators \cite{Bulava:2009ws},
multi-hadron systems\cite{Lat09_Detmold}
and cascade baryons\cite{Lat09_Mathur} 
are reported at the conference.

\bfig
\hspace{-0.05\textwidth}
\begin{minipage}{0.6\textwidth}
\includegraphics[angle=0,width=\textwidth]{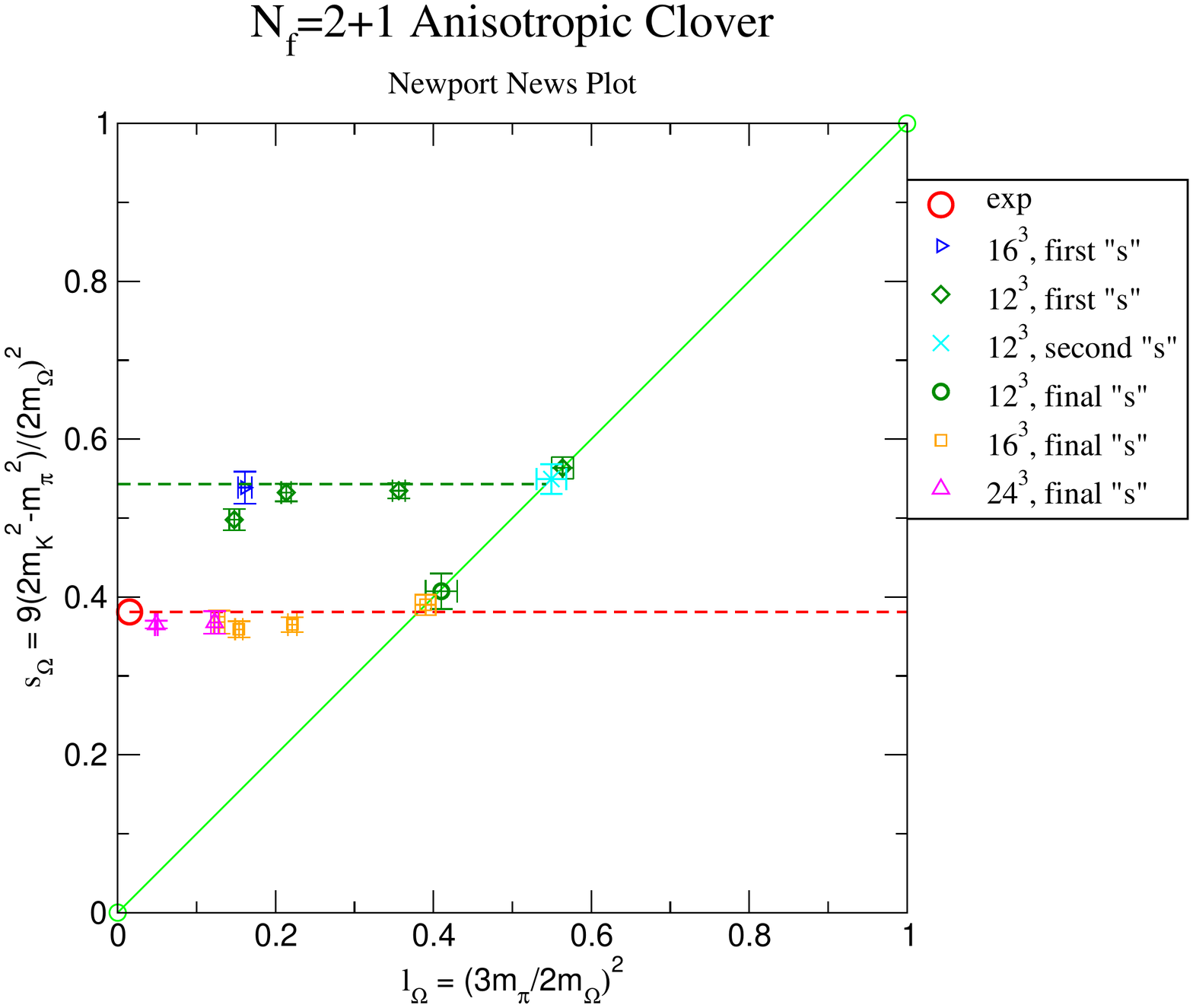}
\end{minipage}
\hspace{-0.05\textwidth}
\begin{minipage}{0.4\textwidth}
\includegraphics[angle=0,width=\textwidth]{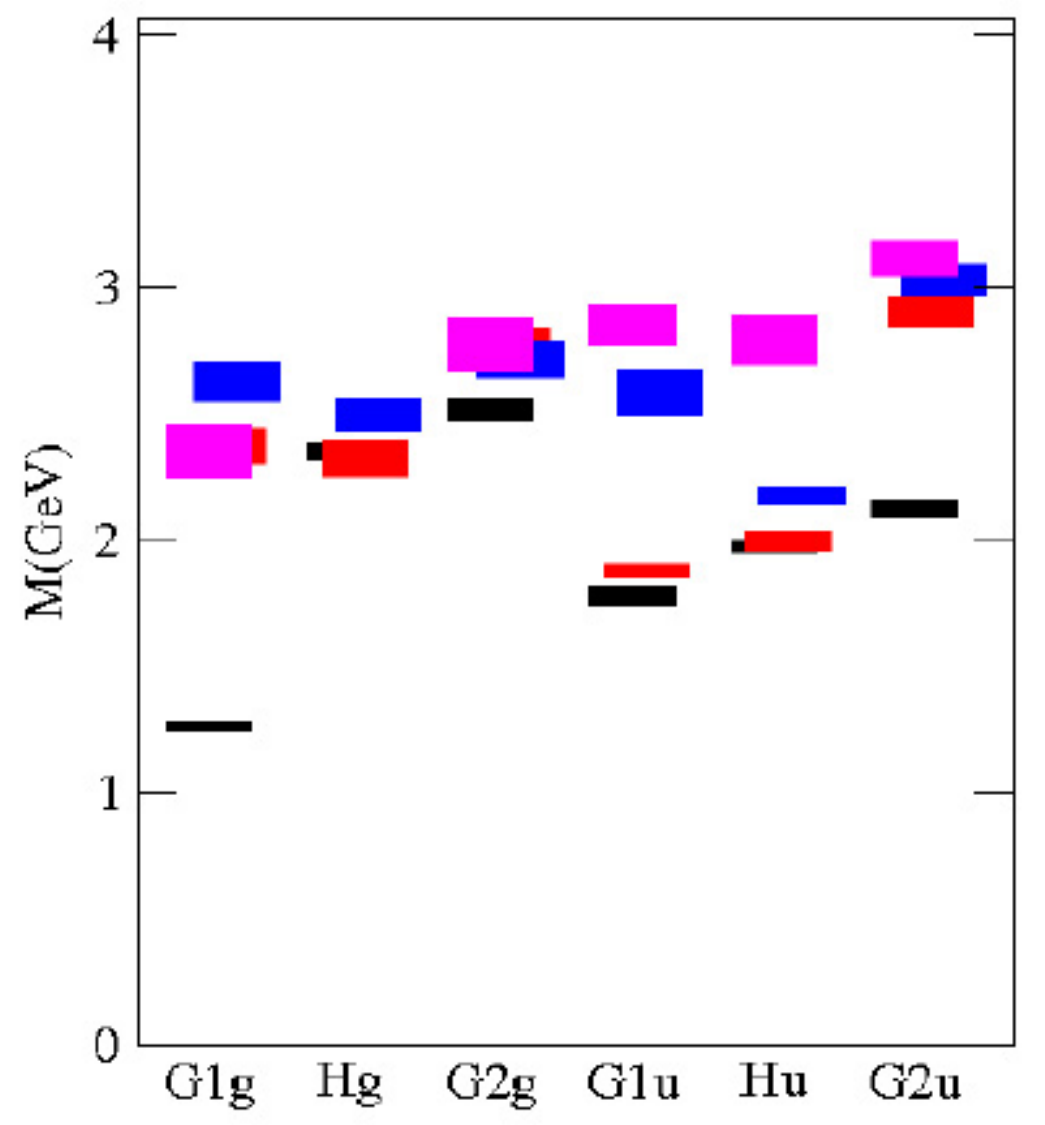}
\end{minipage}
\caption{ $s_\Omega = 9 (2m_K^2-m_\pi^2)/(2m_\Omega)^2$ vs.
$l_\Omega(3m_\pi/3m_\Omega)^2$ (Newport News) plot.}
\label{fig:newport_news}
\caption{ Preliminary results for resonance spectrum
from $N_f=2+1, m_\pi=380$~Mev, $16^3\times 128$ anisotropic
clover ensembles\cite{Cohen:2009zk}. 
}
\efig

\subsection{Overlap and Optimal DWF }
JLQCD  and TWQCD collaboration has been generating $N_f=2$~\cite{Aoki:2008tq}
and $N_f=2+1$ overlap ensembles 
with $a \sim 0.1-0.12$\fm , $L \leq 2$ fm. Recent
results are covered separately in \cite{Lat09_Fukaya,JLQCD:2009sk}. 

Also, TWQCD collaboration has started generating $n_f =$ 2 and 2+1 ensembles 
using optimal domain wall fermion\cite{chiu:2002ir}
with $a^{-1} >\sim$ 1.6~Gev, $L = 2\sim 3$ fm.
These simulations are performed with CUDA-written codes
on Nvidia GPGPU (Nvidia Tesla S1070, 46 Nvidia GTX285/280). 
More details of TWQCD activities can be found in \cite{Lat09_Chiu}.

{
\subsection{Chirally Improved fermion(CI)}
Bern-Graz-Regensburg(BGR) collaboration has started generating $N_f=2$ configurations
using CI action\cite{Gattringer:2008vj}, a 4 dimensional action with links of
length up to 4, tuned to satisfy Ginsparg-Wilson relation approximately. 
Stout smearing is used to suppress discretization error further along  
with L\"uscher-Weisz Gauge action.
\Fig{CI_path} shows collection of links conneced to the point at the center in CI action.
Currently generated ensembles have
$N_f = 2, 16^3\times 32, \sim 2.4$ fm, 318 $< m_\pi < 526$ Mev,
$m_{AWI} = 15 \sim 42$ Mev.
This gives a
significantly different action to be used for studies of excited states to anisotropic Clover
in section \ref{section:AniClo}. 
Preliminary results for excited hadron spectrum is reported in \cite{Lat09_Lang}.
}
\bfig
\begin{minipage}{0.5\textwidth}
\includegraphics[angle=0,width=\textwidth]{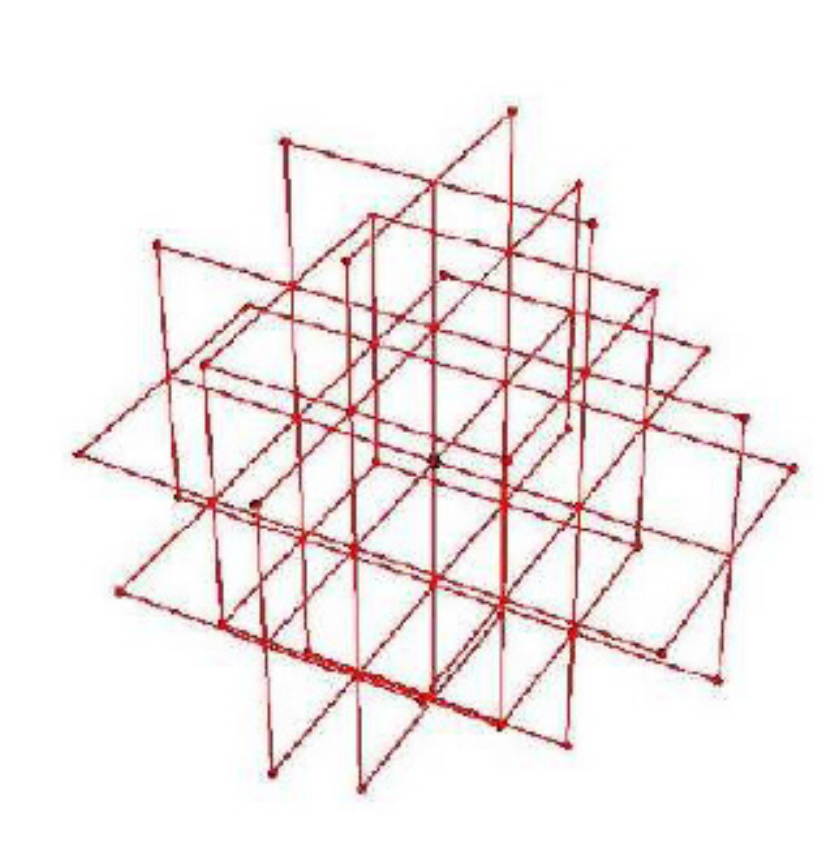} 
\hspace{-0.05\textwidth}
\end{minipage}
\begin{minipage}{0.5\textwidth}
\includegraphics[angle=0,width=\textwidth]{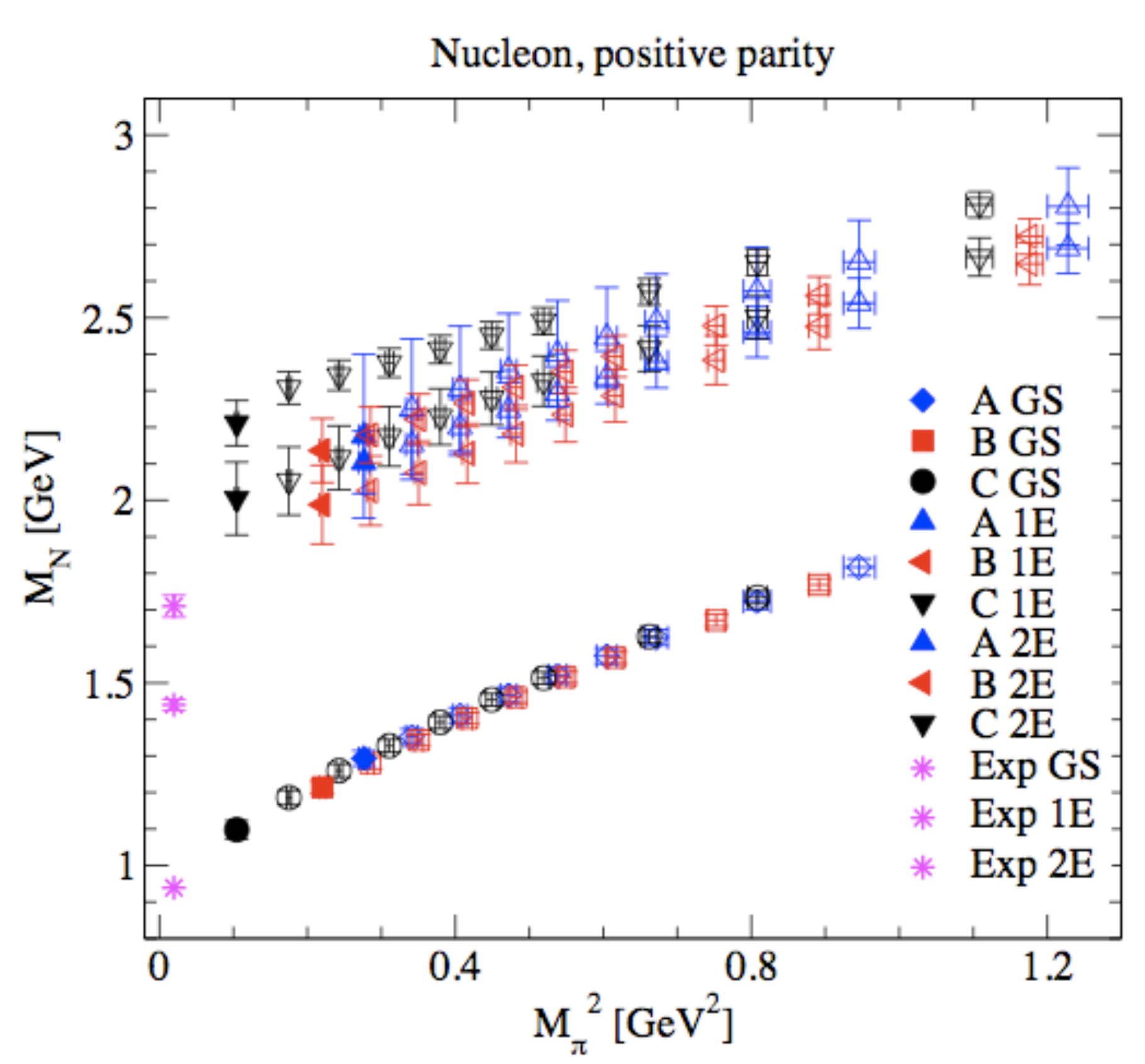} \\
\hspace{0.05\textwidth}
\end{minipage}
\caption{Illustration of paths and Nucleon excited states from CI action, 
from\protect\cite{Lat09_Lang}.}
\label{fig:CI_path}
\label{fig:CI_nuc}
\efig

\vspace{-0.05\textwidth}
\section{Simulation cost}
\label{section:cost}
 
\begin{figure}[hbt]
\hspace{-0.15\textwidth}
\includegraphics[angle=0,width=1.3\textwidth]{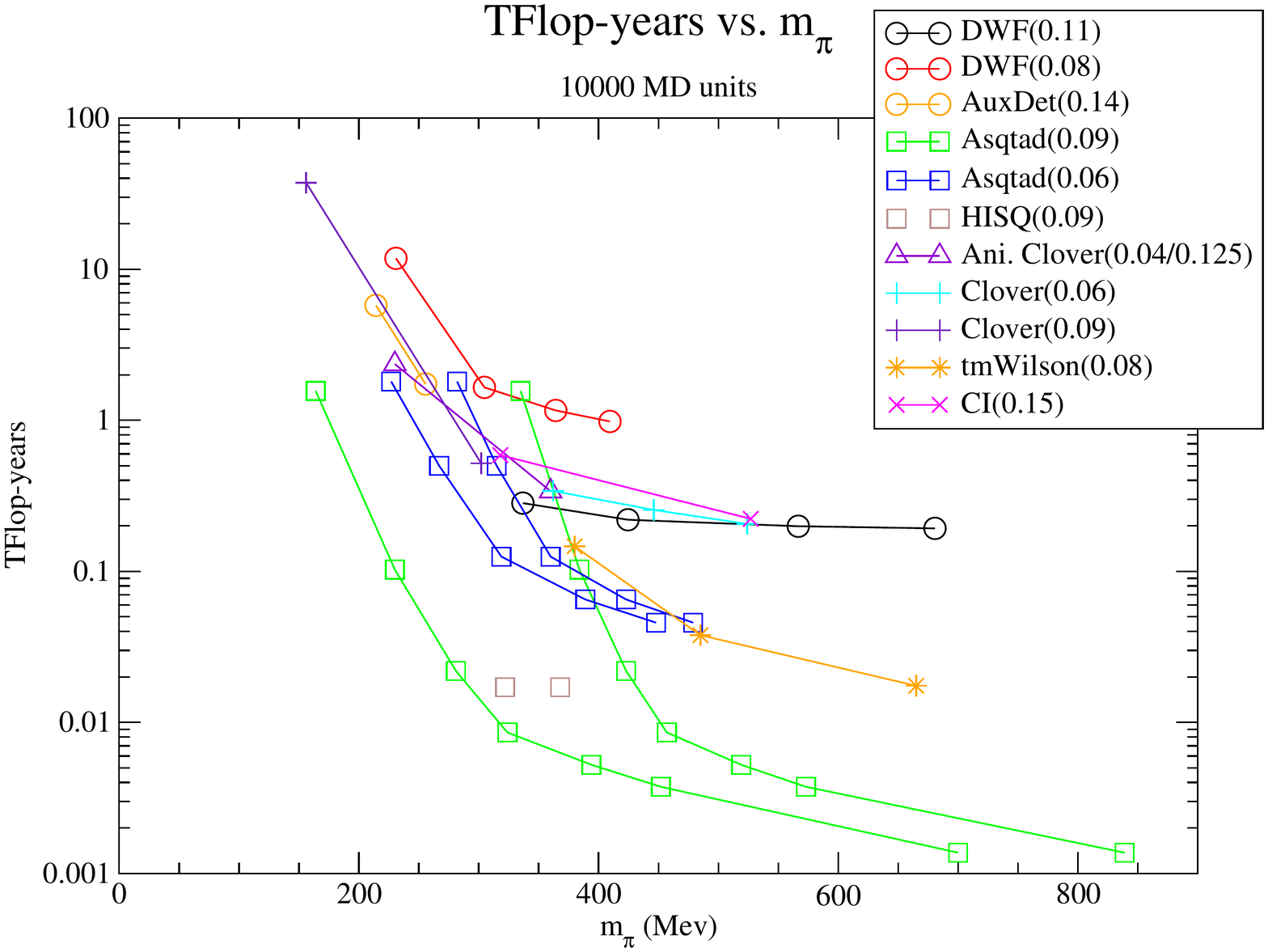}
\vspace{-0.1\textwidth}
\caption{
Total number of TFlop-years needed for $10^4$ MD units versus $m_\pi$(Mev) at
physical volume $L^3 \times T$ satisfying eq.~(\protect\ref{eqn:cost_vol}),
scaled from numbers for existing ensembles by (cost) $ \sim V^{5/4}$.
2 separate lines with same colors for Asqtad and HISQ
ensembles represent the lightest(Goldstone) and the heaviest(taste singlet)
pseudoscalar masses. Numbers for Asqtad and HISQ simulations are calculated
from eq.~\protect\eq{Asqtad_cost}. 
tmWilson(0.08) is from ETMC\cite{Urbach:2005ji}.
Clover(0.09) is from PACS-CS\cite{Aoki:2008sm}. 
Clover(0.006) is from CLS\cite{CLS_cost}.
\label{fig:cost}
}
\end{figure}

As listed in the section above, there are many ongoing dynamical
ensemble generation activities with different discretizations. This is not just because of
increasing availabilty of computing resources, but rather the reflection of
the fact that different actions have different lattice spacing error for
different quantities at same lattice spacings, and the recent progress in ensemble
generation algorithms makes ensemble generation relatively cheaper compared to the
valence propagator generation, either with the same or different discretization, 
and relying on one discretization method to get every physical quantities of interest
may not turn out to be the most cost-efficient approach. This also suggest 
a comparison of different actions at the same lattice spacing is not
necessarily  a fair one.

Having said that, it is still useful to have some idea on how expensive each
action is at a given lattice spacing and quark mass. Here such a comparison is
made by estimating the number of flops to generate the same number of MD
units for each action from available performance data. Recent studies indicate
it is necessary to keep physical volumes to be larger than previously deemed sufficient to control finite size effect, so for a given lattice spacing $a$ and
pseudoscalar meson mass $m_\pi$ for each existing ensembles, the number of
flops are scaled to a volume $L^3\times T$ which  satisfies
\be
L  = \mbox{Max}(4/m_\pi,\quad\ 2.5 \fm),\quad\ T = 2L
\label{eqn:cost_vol}.
\ee

First, available cost formulas
in TFlop-years to generate  10000 MD units or 100 statistically
independent configurations are given below. here
$L, T, a$ are  in fm and $m_{\pi},m_K, m_l, m_s$ are in Mev.
\begin{equation}
Cost[DWF]\cite{Christ:2006zz}: \sim \left(\frac{L^3\times T}{a^4}\right)^{5/4}\left(\frac{1}{m_\pi a}\right) 
\left[ C_0+C_1\frac{1}{m_K^2 a} + C_2 \left(\frac{a}{m_\pi}\right)^2 \right]
\label{eqn:DWF_cost}
\end{equation}

\be
Cost[Asqtad]\cite{MILC_cost}\sim  
\frac{2.9} {64^3\times 144} 
\left(\frac{L^3\times T}{a^4}\right)^{5/4}\frac{0.1}{m_l/m_s}\frac{0.06}{a}
\left(0.5+0.5\frac{0.1}{m_l/m_s}\frac{0.06}{a}\right) 
\label{eqn:Asqtad_cost}
\ee
\begin{gather}
Cost[HISQ] \sim 2 \times Cost[Asqtad] \nonumber 
\end{gather}

\begin{align}
Cost[Clover(DD-HMC)]\cite{DelDebbio:2006cn}
\sim 0.05\times 
\left(\frac{20}{\bar{m_l}}\right)
\left(\frac{L}{3}\right)^5
\left(\frac{0.1}{a}\right)^6
\label{eqn:Wilson_cost}
\end{align}
While these formulas show different dependencies in $a$ or $m_q(m_\pi)$, they all have
the same dependence in volume, (cost) $ \sim V^{5/4}$, which is all we need for the
comparison outlined above.

\Fig{cost} shows the number of total TFlop-years needed
to generate 10000 MD units of gauge configurations with the lattice spacings
and pseudoscalar masses of existing ensembles, with the volume scaled to
satisfy eq.(\ref{eqn:cost_vol}). 
While it should be noted that there are differences in precisely how flops are 
counted for each ensembles which results in some uncertainties in comparing
different ensembles. However,
it does appear that the difference in
total cost between different actions tends to get smaller as the quark mass approaches the physical point.
This is presumably because the low, physical eigenmodes of Dirac operators
increasingly dominate the cost near the chiral limit, while simply the number
of degrees of freedom per sites counts more for simulations at heavier masses. 

Still, the cost is a rapidly changing function of the lattice spacing and
any real cost comparison should be done not only with these numbers,
but in combination with how small lattice spacing error is required for the
quantities one aims to study with each action.

\section{Autocorrelation}
\label{section:autocorrelation} 

\bfig
\bc
\includegraphics[angle=0,width=0.9\textwidth]{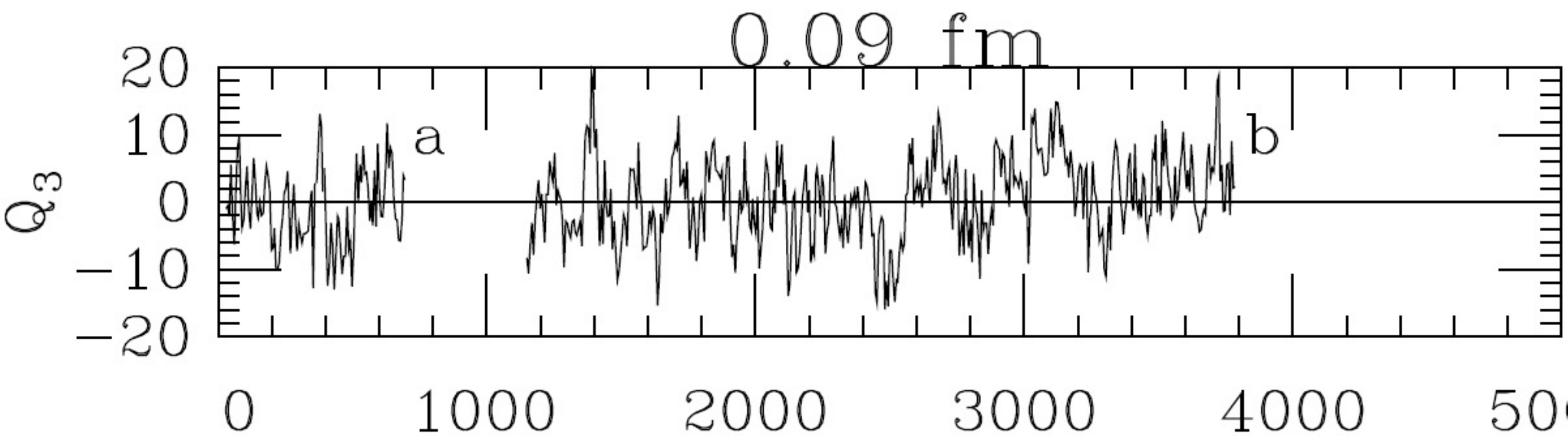}
\includegraphics[angle=0,width=0.9\textwidth]{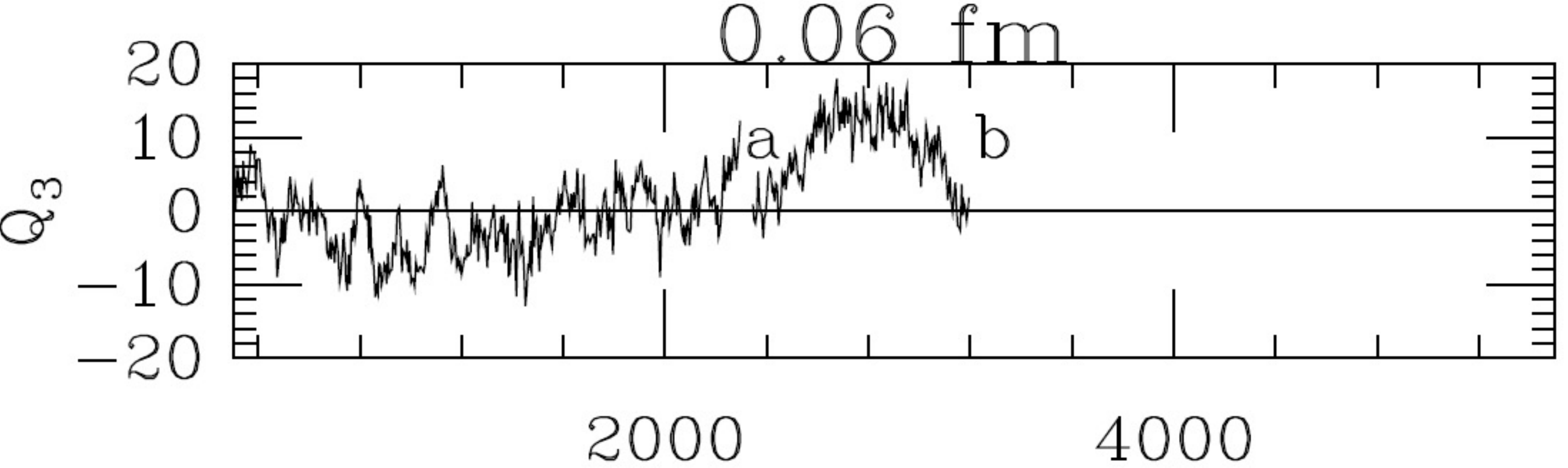}
\includegraphics[angle=0,width=0.9\textwidth]{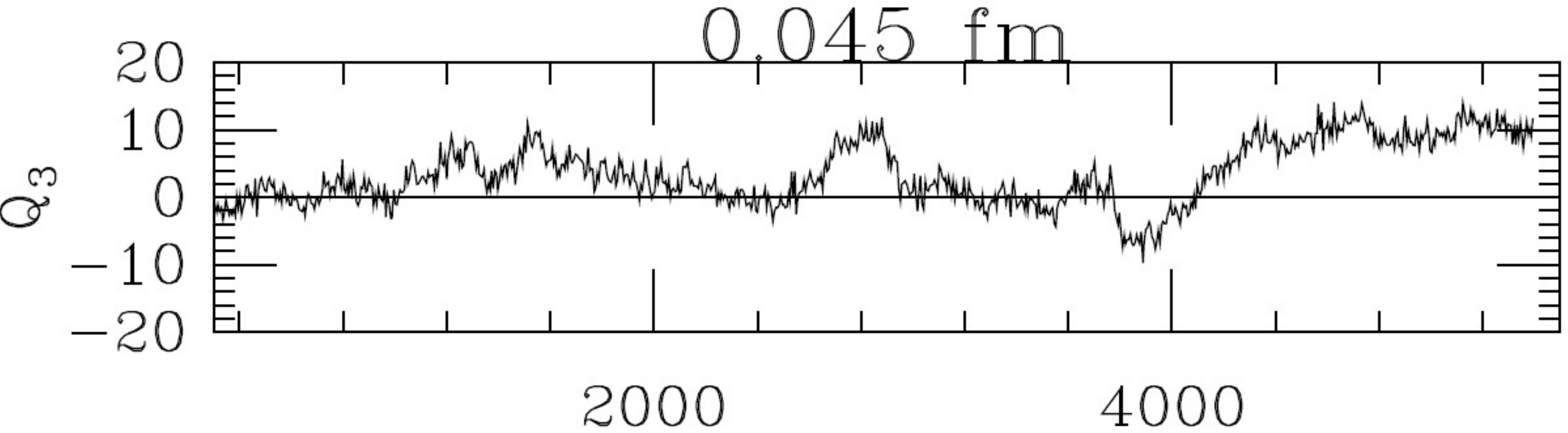}
\ec
\caption{ Time history of global topological charge for Asqtad ensembles at
 $m_l/m_s=0.2~(m_\pi \sim$ 320 Mev)\cite{MILC_top}. The lattice spacings are indicated above each graph. }
\label{fig:asqtad_top}
\efig
\bfig
\vspace{-0.05\textheight}
\bc
\includegraphics[angle=0,width=1.1\textwidth]{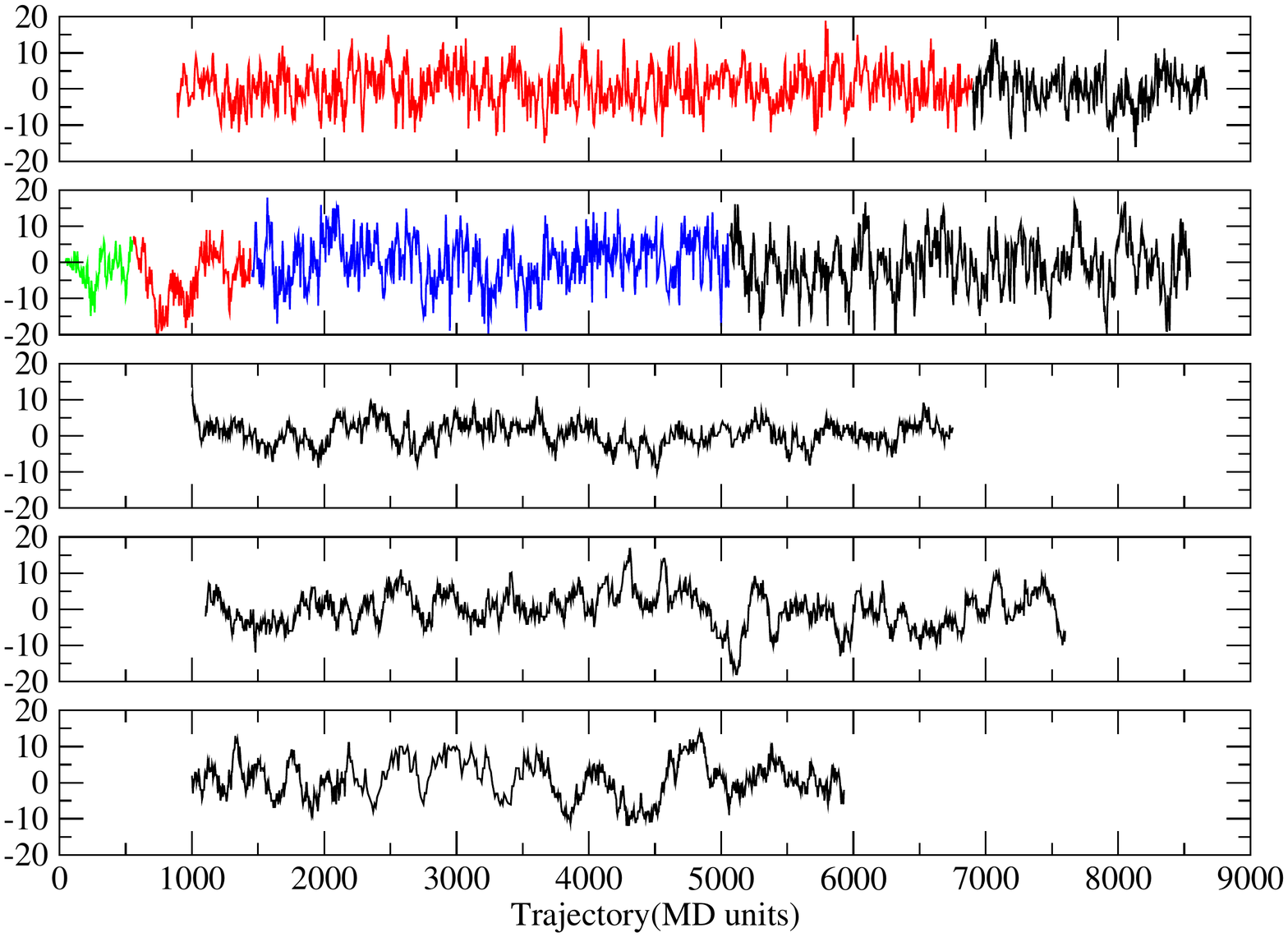} 
\ec
\vspace{-0.05\textheight}
\caption{ Time history of topological charge for DWF ensembles. The ensembles are as
follows from the top: ($a \sim
0.11\fm:  am_l=0.005, 0.01), (a \sim 0.08: am_l = 0.004,0.006,0.008)$.}
\label{fig:DWF_top}
\efig

For obvious reasons, autocorrelations within ensembles is a significant factor
in what the real cost of generating a number of independent configurations is. 
Unfortunately, in practice this is
not necessarily something one can measure easily,
even after ensemble generation is nominally finished, as it can and does vary greatly between different quantities.

While all the ensembles reported here have reasonable ($<100$ MD units)
autocorrelation time for quantities such as plaquette, meson propagators and
number of CG iterations, many of them only a few MD units,
 a significant slowdown of change in global topological charge was observed in ensembles
with relatively smaller lattice spacings. Time evolution of topological charge  
in both asqtad and DWF ensembles are shown in
Figures \ref{fig:asqtad_top} and \ref{fig:DWF_top}. 
A similar behavior was also observed in clover action simulations with DD-HMC\cite{Schaefer:2009xx}. 
While the global topological charge itself may or may not be relevant in physics
as long as the physical volume is large enough, 
it still is a sign of failures of evolution algorithms currently in use to
ensure ergodicity of each ensemble.

While these results are still preliminary and requires  more careful studies, if not
longer ensembles, it would
be crucial to check this for any ensembles, especially
ensembles with smaller lattice spacings, since these ensembles are designed to
have systematic errors small enough for precision studies, and long
autocorrelations affect the cost of generating independent configurations to
control the statistical errors significantly, or introduce systematic errors
from failing to sample all the topological sectors.

This also presents a potential problem in relying on staying on the same lattice
discretization and increasingly smaller lattice spacing, made possible by
increasing computing resources, to control the discretization error. 
This might make it preferable to make better use of increasing computing
resources by keeping the lattice spacing moderate and try more improved
actions with smaller lattice spacing error,
instead of continuing more conventional actions at smaller lattice spacing.
Unfortunately, currently this approach - comparing different discretizations - is hampered
substantially by the lack of a quantity easily measurable with known and
accurate physical value to fix the lattice spacing, as shown by different
values of $r_{0,1}$'s used or measured for different discretizations in
section~\ref{section:action}.

This is not to say this problem cannot be circumvented by algorithmic
development. The fact that the increase in autocorrelation is largely
independent of the quark mass, choice of discretization and simulation
algorithms\cite{Schaefer:2009xx}} suggests the gauge action is the culprit, and techniques for more aggressive
decorrelation while keeping the action the same, such as Noisy Monte
Carlo\cite{Bakeyev:2000wm}, 
or even drastically different ideas such as
\cite{Luscher:2009eq},
could turn out to be effective.

\section{Algorithms and  techniques}
\label{section:algorithm} 
In the last several years, there have been much progress in fermion simulation 
algorithm. Improvements such as Rational Hybrid Monte Carlo\cite{Clark:2006fx}, higher order
 integrators, particularly Omelyan\cite{Takaishi:2005tz,Omelyan}, multiscale
integrators  and mass
preconditioning\cite{Hasenbusch:2002ai} achieved 
almost ubiquitous usage, similar to what $\Phi$ and R algorithm\cite{Gottlieb:1987mq} used to be.
While these advances made details of fermion action unique for each simulation and made
extensive tuning necessary, they also made, in combination with advances in hardware, 
the dynamical simulations with pion masses at or close physical value
 and made the cost of ensemble generation much less dominating compared to valence
propagator generation needed for various analyses.
This is certainly not to say we have exhausted the possibility of improvement.
Algorithmic development such as force gradient integrator and 
better tuning via shadow hamiltonian\cite{Kennedy:2009fe,Chin:2000zz} suggests
further reduction in cost of realistic QCD ensemble generation is possible.

Here some of more recent advances in algorithms are summarized, starting with
mass reweighting technique, which has shown to be surprisingly useful.

\subsection{Reweighting}
\label{section:reweighting}
The basic Idea of reweighting is well known. There are more than one
Hamiltionians used in typical Monte Carlo simulations:
\bi
\item Guiding Hamiltonian for Molecular Dynamics:\\
 $ dU/dt_{MD} = -i H U, \quad\ d {\cal H}_1/dt_{MD} = 0 $
(As pointed out, for example in \cite{Kennedy:2009fe}, MD integrators are not exact. Here
$\H_1$ is the hamiltonian used to construct MD integrator, not the one actually being
preserved by it .)
\item Hamiltonian for Metropolis step :\\
Acceptance =
Min$\left(1,\exp\left[-(\H_2({U}_{i+1})-\H_2({U}_{i}))\right]\right)$
\item Hamiltonian for ensemble averaging 
\be
< O >_{\H_3} = \frac{\int [dU] O(U) W(U)}{\int [dU]W(U)},\quad\ W(U) =
\exp{\left[-(\H_3(U)-\H_2(U))\right] }
\label{eqn:rw_fac}
\ee
\ei
Typically it is called "reweighting" when $\H_2$ is intentionally set
different from $\H_3$.
In principle $\H_1,\H_2,\H_3$ can be all different and there are working examples,
ranging from subtle ones such as 
using less  precision and/or relaxed stopping condition for Hybrid Monte Carlo, to more
explicit ones used to avoid singularities in HISQ action(section \ref{section:HISQ}) where
$\H_1 \neq \H_2$, to
many thermodynamic studies, to locate the phase transition 
temperature\cite{Cheng:2006qk} or circumvent difficulties with finite density\cite{Fodor:2004nz} where $\H_2 \neq \H_3$.  

However, $\H_{1,2,3}$ are extensive quantities ($\H \propto V$) and conventional wisdom has been that it
is  hard to find $\H$'s which the acceptance/reweighting factor $e^{-\Delta \H }$ is close
enough to 1 while it is different enough to give significant benefit for simulations with
larger volumes. (For example, $32^3$ DWF simulations reported in section
\ref{section:DWF} has $\H$ on the order of $10^8$.) 

However, recently some more working examples have been reported, 
such as reweighting of
light quark \cite{Hasenfratz:2008fg,Luscher:2008tw} and
strang quark \cite{Tsukuba_Jung,Aoki:2009ix} toward the physical point.
While more aggressive reweighting such as those for light quark have potential to be more
beneficial in the long run, the reweighting of strange quark mass in
particualr has proved to be a cost effective way of eliminating one of the major sources
of systematic error and provide an immediate benefit.  
A more detailed description of strange quark reweighting is given below.

\subsubsection{Reweighting of dynamical strange quark }
\label{section:str_rw}

Due to the nonperturbative nature of QCD, lattice spacing of any lattice QCD simulation,
with or without fermions,  is not known {\it a priori }
until it is measured on thermalized configurations. 

Somewhat ironically, this has not posed problem for light quarks in
practice as much, as
typically multiple ensembles of different light quark masses are generated to do extrapolations to
the chiral limit anyways, recent simulations near or at physical point
notwithstanding. However, it is in principle possible and would be quite beneficial
to simulate at the correct physical strange quark.
Unfortunately, this is not the case and the dynamical strange quark is often
different from physical value by up to 20\%. 
Traditionally one of these approaches has been taken to address this problem:

\bi
\item Generate multiple ensembles with different dynamical strange quark masses near
physical value and use interpolation
\item  Using SU(3) ChPT to fit up to the strange quark.
\item Do multiple parameter tuning runs in smaller volume before larger runs.
\item Include the effect of discrepancy as a systematic error.
\ei

None of these approaches are particularly attractive and can be time consuming.
It can save significant computing (and human) resources if this can be avoided.

It turned out the reweighting factor $W(U)$ (Eq.(\eq{rw_fac})) is often close
to 1 even when strange quark mass is $10 \sim 20 \%$ different from the
dynamical value, which makes the tuning of strange quark to the physical value
via reweighting possible. 
This technique can also be used
to calculate derivative with respect to the dynamical quark mass\cite{Ohki:2009mt}.

Reweighting factor for 1 flavor can be calculated as follows:
\begin{gather}
W([U],m_s,m_s') = \det\left(\frac{D'^{\dagger}D'}{D^{\dagger}D}\right)^{1/2} 
= \mbox{det}(\Omega)^{-1/2}, 
\Omega([U],m_s,m_s') = D'^{-1}DD^\dagger(D'^\dagger)^{-1} \nonumber \\
D = D([U],m_l,m_s), D' = ([U],m_l,m_s') \nonumber \\
W([U],m_s,m_s') = \frac{\int d\xi d\xi^\dagger e^{-\xi\dagger
\sqrt{\Omega([U],m_s,m_s')} \xi} } {\int d\xi d\xi^\dagger e^{-\xi\dagger \xi}}  = 
\left<e^{-\xi^\dagger(\sqrt{\Omega([U],m_s,m_s')} -1)\xi} \right>\label{eqn:rat_quo}
\end{gather}
where $\xi$ is a Gaussian random vector. $\sqrt{\Omega[U]}$ can be calculated using Rational approximation, similar to RHMC.
Alternatively, the same reweighting factor can be calculated by
\be
W([U],m_s,m_s')= \sqrt{\left<e^{-\xi^\dagger(\Omega[U] -1)\xi} \right>}\label{eqn:quo}
\ee
and it could be more efficient, as \eq{quo} can be calculated with just one
inversion per random number.
However, \eq{quo} is in principle a biased estimator and more
accuracy than \eq{rat_quo} is required for reweighting factors for each configurations
to ensure the effect from inaccurate reweighting factor is under control, while
\eq{rat_quo} can be used without such requirement.
Another scheme to use Taylor expansion to calculate square root is used in
\cite{Aoki:2009ix}.

Due to the nature of \eq{rat_quo} where each evaluation using random vector is
exponentiated before average, it is essential to ensure $M = \sqrt{\Omega([U],m_s,m_s')}$ has
only eigenvalues close to 1. In fact, it can be shown that the error in \eq{rat_quo} 
diverges if $M$ has eigenvalues smaller than 1/2. Even when this is not
the case, eigenvalues significantly away from 1 can make the reweighting factor
converge very slowly with finite statistics.
For mass reweighting, determinant breakup by intermediate masses\cite{Hasenfratz:2008fg}
\begin{gather*}
W([U],m_s,m_s') = \Pi_{i=0}^{n_m-1} \left<e^{-\xi^\dagger(\sqrt{\Omega([U],m_i,m_{i+1})}
-1)\xi} \right>, \\
m_s = m_0<m_1<\cdots m_{n-1}<m_n=m_s' \mbox{or vice versa}.
\end{gather*}
gives an additional benefit that the reweighting factors for the masses between $m_s$ and
$m_s'$ is automatically available. For other reweighting where this is not
available, for example~\cite{Lat09_Tomomi}, Breakup using rational approximation, similar to "root-N
trick" in \cite{Clark:2006fx} is applicable.

Now, observables for $m_s'$ is calculated from ensemble $[U_i]$ generated $m_s$ by 
\begin{equation}
\left< O \right> (m_s') = \frac{\Sigma_i O[U_i] W([U_i],m_s,m_s')}{\Sigma_i w[U_i]}
\end{equation}

\hspace{-0.05\textwidth}
\bfig
\vspace{-0.12\textheight}
\begin{minipage}{0.5\textwidth}
\includegraphics[angle=0,width=1.1\textwidth]{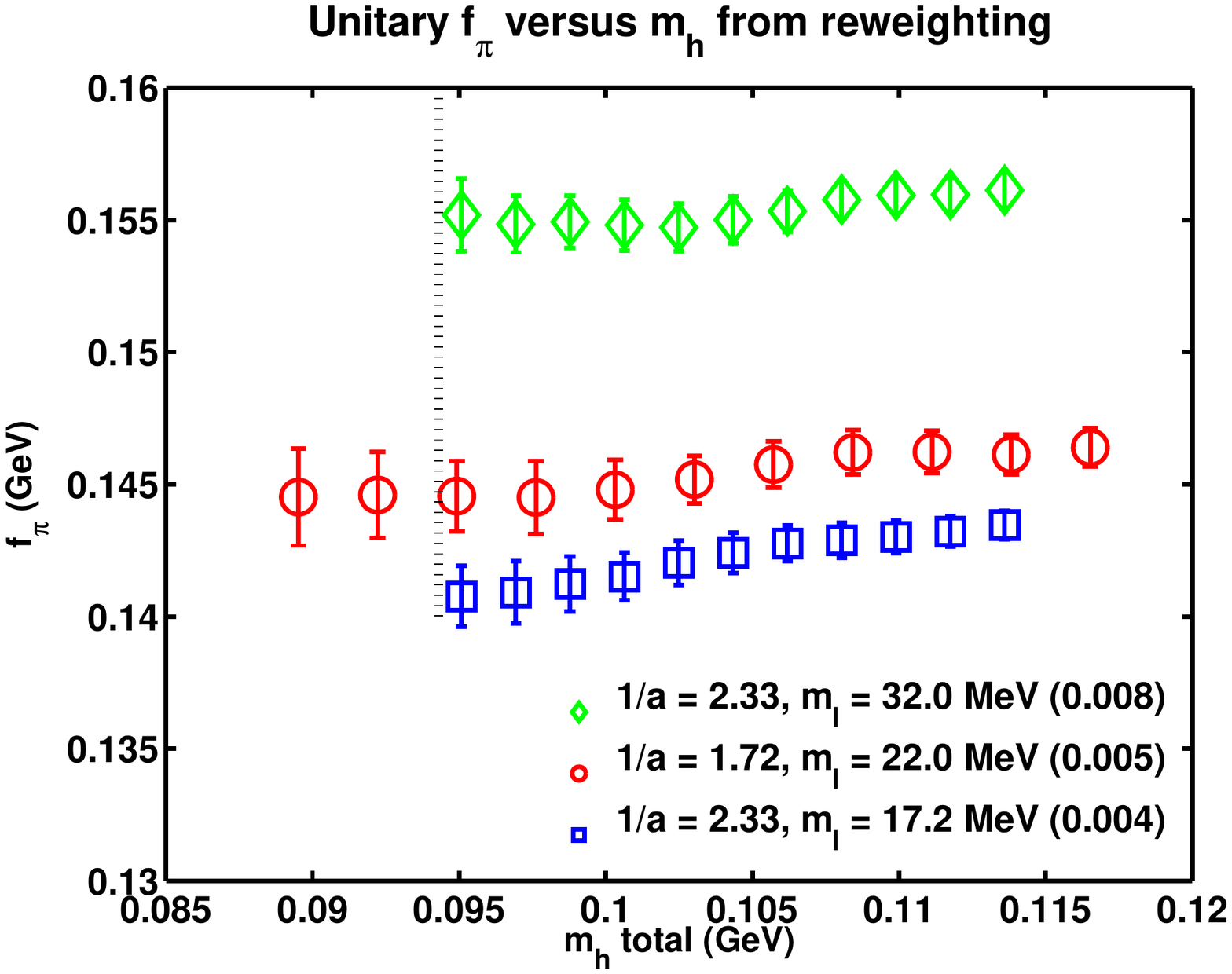}
\end{minipage}
\begin{minipage}{0.5\textwidth}
\includegraphics[angle=0,width=1.1\textwidth]{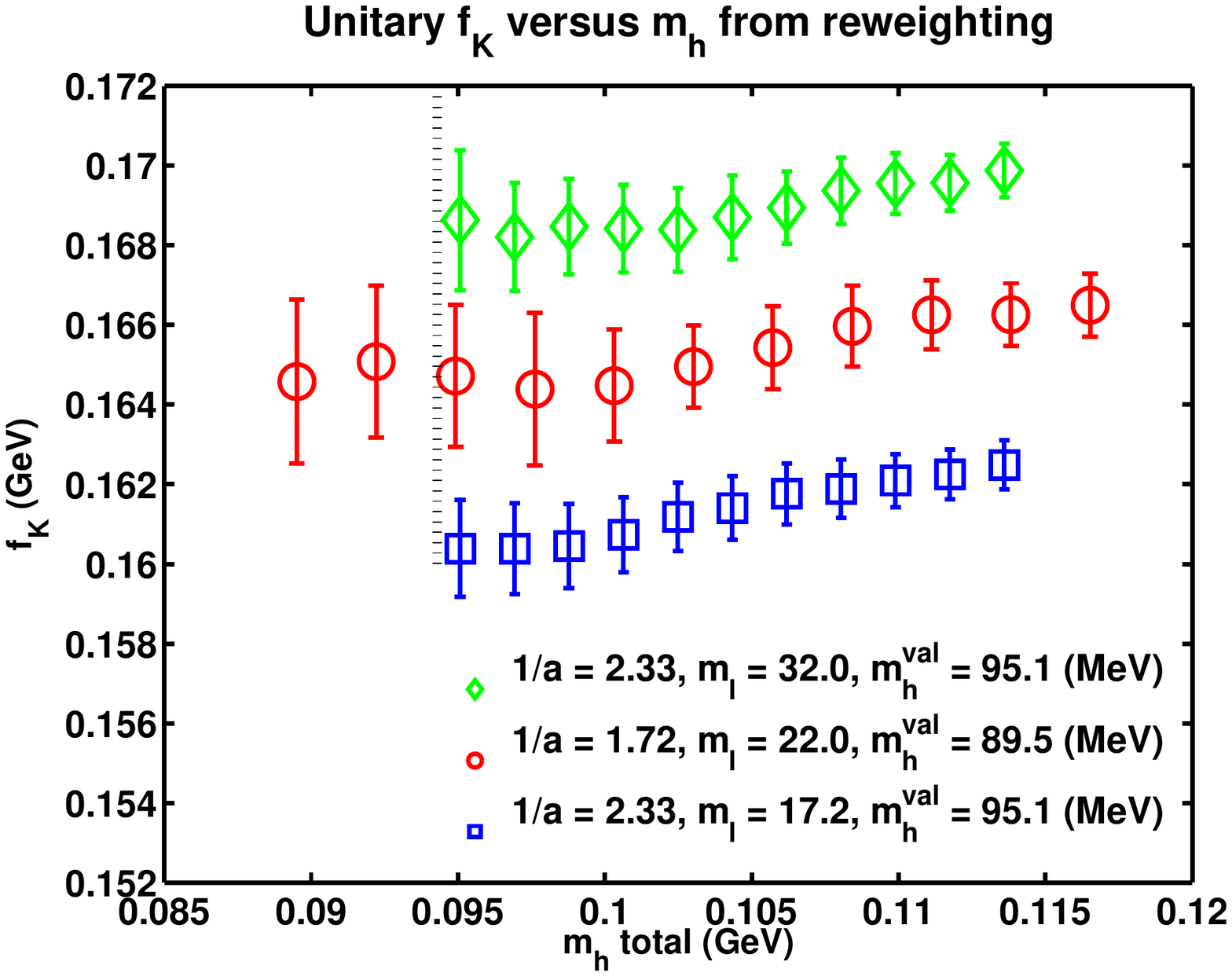}
\end{minipage}
\vspace{-0.1\textheight}
\caption{ Reweighted $f_\pi$ and $f_K$ for $N_f=2+1$ DWF ensembles, from\cite{Mawhinney:2009jy}. 
Broken vertical lines represent the physical strange quark mass and the rightmost points shows the original
sea strange quark mass for each ensemble.} 
\label{fig:DWF_rw}
\efig

Figure \ref{fig:DWF_rw} shows reweighted pseudoscalar decay constant of DWF ensemble in
section~\ref{section:DWF}. Even though only 4 random vectors per mass step is
used to calculate reweighting factor for $a^{-1} \sim$ 2.33~Gev, $a m_l = 0.004$
ensemble, the reweighted values at $\sim 20 \%$ smaller than $m_s$ has only slightly
larger error than the error at $m_s$.
As shown in section \ref{section:Clover}, PACS-CS collaborations has done both strange
quark and light quark reweighting to reach physical point\cite{Aoki:2009ix} and 
tuning of twist for the strange/charm quarks via reweighting is also being
carried out by ETMC collaboration\cite{Baron:2008xa}.

Given the success of mass reweighting,
we can also think of other cases where reweighting can be helpful.
\bi
\item $\H_1 \neq \H_2$:
\bi
\item Using a relaxed approximation sign function during MD step of
Overlap/Domain wall fermions, to
avoid topology tunneling difficulties while preserving good chiral symmetry, this
appears to be promising if the phase space where $\H_1 \neq \H_2 \geq 1$ is limited.
One approach using DWF, namely
reweighting to larger $L_s$ in DWF was explored in \cite{Lat09_Tomomi}.
\ei
\item $\H_2 \neq \H_3$:
\bi
\item Apply reweighing to part or all of differences in action in mixed action studies.
This would effectively trade systematic error with a statistical one.
\item Adding new terms to the action via reweighting,
 such as adding charm quarks to
$N_f=2+1$ configurations or adding QED for studies of electromagnetic
effects\cite{Duncan:2004ys}.
\ei
\ei

\subsection{Nested/mixed precision solvers}

Advent of many multicore chips, available currently or in the near future,
 present in principle significantly more computing 
power than currently available. However, this increase in computing power often comes without 
corresponding increase in bandwidth, both in   
memory  and network. 
While it is not yet clearly shown that these architecture can achieve scalability suitable for large scale dynamical ensemble generations using conventional algorithms such as (R) HMC, or newer algorithms such as DD-HMC, it still is expected to
provide competitive hardware platforms for propagator generation, where
``embarrassingly parallel'' -
the local volume is chosen to be optimal for performance and run multiple jobs
concurrently - approach is applicable.

The relative lack of memory bandwidth also pose problems for propagator
generation even when it is possible to fit it on one node to circumvent the
network bottleneck. One way to lessen the pressure on memory bandwidth is to use
less precision arithmetic. While the relative
difference in speed between single and double precision is significant for
many existing platforms, 
until recently most frequently used  approaches to take advantage of this was 
somewhat simple-minded, such as
first solving in single precision and finish with double(less effective if
small residual is required),
MD in lower precision, Metropolis in double precision (if $\Delta H$ from
precision loss is small, expected to scale poorly for larger volume), or even
using single precision MD and skip Metropolis when the step size error is deemed smaller than other systematic errors.

Recently there have been rediscovery of algorithms, such as
defect correction(used in \cite{Durr:2008rw}) and  Reliable
updates\cite{RelUpdate} for effective use of lower precision solvers for
double precision inversions.
Reliable update technique in particular has shown to be very effective in using single or half(16bit)
precision, useful in  GPGPU.  More detailed description can be found in
\cite{Lat09_Clark,Clark:2009wm}.
In addition to this, there are other nested inversion algorithms such as
Generalized Conjugate Residual(GCR),  
used with Domain decomposition\cite{Luscher:2003qa} or inexact
deflation\cite{Luscher:2007se}. A more extensive studies of these algorithms 
could allow much more effective inversion algorithms on existing and emerging architectures.

\section{Conclusions}
Recent advances in dynamical fermion simulation algorithms and continuing progress in
hardware has enabled multiple collaborations to generate dynamical
ensembles with different lattice discretizations, lattice spacings and dynamical
quark masses.
This is a striking departure from only a few years ago, when Asqtad configurations
generated by the MILC collaboration were the only choice for continuum extrapolation with
multiple lattice spacings and quark masses. Some of the ensembles are even being
generated at or close to the physical pion mass.
$N_f=2, 2+1 $ and 2+1+1 dynamical lattice QCD ensembles generation activities of
various collaborations are summarized. It is shown that 
computing resources on the order of 10 to 100
TFlop-years, well within reach with technologies currently available, is
enough to generate $N_f=2+1$ ensembles with $>100$ independent configurations
at $m_\pi <200$ Mev and $m_\pi L >4$, barring the effect of autocorrelation
observed by global topological charge. 

A steep dependence of simulation cost on the lattice spacing and
different scaling behavior of different actions, as well as the ambiguities in
defining the lattice spacing itself, makes superficial comparisons 
between different actions not necessarily useful. 
Although it has become possible to use
relatively inexpensive actions such as Wilson to generate ensembles very close
to or at physical quark mass, the presence of significant
autocorrelations observed by topological charge evolution at smaller lattice
spacings ($a \leq 0.08$fm) should be better understood and circumvented to ensure usefulness
of large scale ensembles.
It is possible that multiple, more highly improved actions at a moderate
lattice spacing is a better way to ensure that systematic errors are under control for
all the quantities we are interested in studying.
Presence  of many ensemble generation activities is helpful in that respect.

Some of the recent progress in algorithms and techniques were also summarized. The mass
reweighting technique, which turned out to be quite effective in eliminating one of the
persistent source of systematic error is explained in more detail.  
  Reweighting in general could alleviate the need to generate separate
ensembles  for slightly different parameters 
and/or actions, even for relatively large lattices.

\vspace{0.02\textheight}
\noindent
{\bf Acknowledgments}
\vspace{0.01\textheight}

I would like to thank the colleagues who sent previously unpublished materials
before the conference for preparation of the talk, in particular 
A. Bazavov, T. Blum, T-W. Chiu, C. Detar, S. Gottlieb, G. Herdoiza, R.
Horsley, B. Joo, Y. Kuramashi, C. B. Lang, R. D. Mawhinney, D. Palao, S.
Reker, G. Schierholz and H. Wittig. I am also indebted to many RBC/UKQCD
colleagues for many helpful discussions. 
The author was supported by the U.S. DOE under contract DE-AC02-98CH10886.

\begin{spacing}{0.8}
\providecommand{\href}[2]{#2}\begingroup\raggedright\endgroup
\end{spacing}

\end{document}